\documentclass{aa}  

\usepackage{soul}
\usepackage{graphicx}
\usepackage{subfigure}
\usepackage{txfonts}
\usepackage[]{hyperref}
\hypersetup{colorlinks=true,citecolor=blue}
\begin{document}

   \title{Imaging the Molecular Interstellar Medium in a Gravitationally Lensed Star-forming Galaxy at z=5.7}

   \author{Yordanka Apostolovski \inst{1} \fnmsep \inst{2}
            \and Manuel Aravena \inst{3}
            \and Timo Anguita \inst{1} \fnmsep \inst{2}
            \and Justin Spilker \inst{4}
            \and Axel Wei\ss \inst{5}
            \and Matthieu B\'ethermin \inst{6}
            \and Scott C. Chapman \inst{7}
            \and Chian-Chou Chen \inst{8}
            \and Daniel Cunningham \inst{7} \fnmsep \inst{9}
            \and Carlos De Breuck \inst{8}
            \and Chenxing Dong \inst{10}
            \and Christopher C. Hayward \inst{11}
            \and Yashar Hezaveh \inst{11} \fnmsep \inst{18}
            \and Sreevani Jarugula \inst{13}
            \and Katrina Litke \inst{14}
            \and Jingzhe Ma \inst{15}
            \and Daniel P. Marrone \inst{14}
            \and Desika Narayanan \inst{10} \fnmsep \inst{16} \fnmsep \inst{17}
            \and Cassie A. Reuter \inst{13}
            \and Kaja Rotermund \inst{7}
            \and Joaquin Vieira \inst{13}
          }

   \institute{{Departamento de Ciencias Fisicas, Universidad Andres Bello, Fernandez Concha 700, Las Condes, Santiago, Chile} \\
   \email{yordanka.apostolovski@gmail.com}
    \and{Millennium Institute of Astrophysics (MAS), Nuncio Monse\~nor Sotero Sanz 100, Providencia, Santiago, Chile}
    \and {N\'ucleo de Astronom\'ia, Facultad de Ingenier\'ia y Ciencias, Universidad Diego Portales, Av. Ejercito 441, Santiago, Chile}
    \and{Department of Astronomy, University of Texas at Austin, 2515 Speedway Stop C1400, Austin, TX 78712, USA}
    \and{Max-Planck-Institut f\"{u}r Radioastronomie, Auf dem H\"{u}gel 69, D-53121, Bonn, Germany}
    \and{Aix Marseille Univ., Centre National de la Recherche Scientifique, Laboratoire d’Astrophysique de Marseille, Marseille, France}
    \and{Department of Physics and Atmospheric Science, Dalhousie University, Halifax, NS, B3H 4R2, Canada}
    \and{European Southern Observatory, Karl Schwarzschild Stra\ss e 2, 85748 Garching bei M\"{u}nchen, Germany}
    \and{Department of Astronomy and Physics, Saint Mary's University, Halifax, NS, B3H 3C3, Canada}
    \and{Department of Astronomy, University of Florida, Gainesville, FL 32611, USA}
    \and{Center for Computational Astrophysics, Flatiron Institute, 162 Fifth Avenue, New York, NY 10010, USA}
    \and{Kavli Institute for Particle Astrophysics and Cosmology, Stanford University, Stanford, CA 94305, USA}
    \and{Department of Astronomy, University of Illinois, 1002 West Green St., Urbana, IL 61801, USA}
    \and{Steward Observatory, University of Arizona, 933 North Cherry Avenue, Tucson, AZ 85721, USA}
    \and{Department of Physics \& Astronomy, University of California, Irvine, CA 92697, USA }
    \and{University of Florida Informatics Institute, 432 Newell Drive, CISE Bldg\ E251, Gainesville, FL 32611, USA}
    \and{Cosmic Dawn Center at the Niels Bohr Institute, University of Copenhagen\
 and DTU-Space, Technical University of Denmark}
    \and{Department of Physics, University of Montreal, Pavillon Roger-Gaudry (D-428) 2900 boul. Édouard-Montpetit, Canada }
             }

   \date{Received XXX ; accepted YYY}

 
  \abstract
   {}
   {We present and study spatially resolved imaging obtained with the Atacama Large Millimeter/submillimeter Array (ALMA) of multiple $^{12}$CO($J=$6$-$5, 8$-$7 and 9$-$8) and two H$_2$O(2$_{02}-$1$_{11}$ and 2$_{11}-$2$_{02}$) emission lines and cold dust continuum toward the gravitationally lensed dusty star forming galaxy SPT0346-52 at z=$5.656$. }
   {Using a visibility-domain source-plane reconstruction we probe the structure and dynamics of the different components of the interstellar medium (ISM) in this galaxy down to scales of 1 kpc in the source plane.}
   {Measurements of the intrinsic sizes of the different CO emission lines indicate that the higher J transitions trace more compact regions in the galaxy. Similarly, we find smaller dust continuum intrinsic sizes with decreasing wavelength, based on observations at rest-frame 130, 300 and 450$\mu$m. The source shows significant velocity structure, and clear asymmetry where an elongated structure is observed in the source plane with significant variations in their reconstructed sizes. This could be attributed to a compact merger or turbulent disk rotation. The differences in velocity structure through the different line tracers, however, hint at the former scenario in agreement with previous [CII] line imaging results. Measurements of the CO line ratios and magnifications yield significant variations as a function of velocity, suggesting that modeling of the ISM using integrated values could be misinterpreted. Modeling of the ISM in SPT0346-52 based on delensed fluxes indicate a highly dense and warm medium, qualitatively similar to that observed in high redshift quasar hosts.}
   {}



   \titlerunning{Imaging the ISM in a starburst at z=5.7}
   \authorrunning{Y. Apostolovski et al.}
   \maketitle
%

\section{Introduction} \label{sec:intro}

\begin{table*}
\caption{Observational summary of data}              
\label{tab:summary}      
\begin{tabular}{ccccccccc}          
\hline\hline                        
Tel & Line & $\nu_{\text{obs}}$ & $\nu_{\text{rest}}$ & Dates & Time On-Source & Beam Size \tablefootmark{a} & PA & $\sigma_{\text{rms}}$ \tablefootmark{b}  \\
 & & (GHz) & (GHz) & (MMDDYY) & (hours) &  & (deg) &($\mu$Jy)
 \\    
\hline                                   
ALMA & CO(6$-$5) & $103.88$ & $691.47$ & $042816$ & $0.57$ & $1\arcsec.2 \times 1\arcsec.1$ & $-11.06$ & $75$ \\
ALMA & CO(8$-$7) & $138.49$ & $921.79$ & $042315, 042715, 050315$ & $0.86$	& $2\arcsec.6 \times 1\arcsec.5$ & $88.08$ & $130$ \\
ALMA & CO(9$-$8) & $155.78$ & $1036.91$ & $042315$ & $0.31$ & $2\arcsec.0 \times 1\arcsec.3$ & $88.54$ & $190$ \\
APEX \tablefootmark{$\dagger$} & CO(10$-$9) & $173.08$ & $1151.98$ & $071517$ & $34.1$ & $-$ & $-$ & $-$ \\
ALMA & H$_2$O$(2_{11}-2_{02})$ & $112.99$ & $752.03$ & $041616$ & $0.57$ & $1\arcsec.5 \times 1\arcsec.4$ & $79.69$ & $235$ \\
ALMA & H$_2$O$(2_{02}-1_{11})$ & $148.42$ & $987.92$ & $042315, 042715, 050315$ & $0.86$ & $2\arcsec.5 \times 1\arcsec.5$ & $-80.98$ & $240$ \\
\hline                                             
\end{tabular}
\tablefoottext{a}{Beam size for a Briggs weighted image with a robustness parameter of $0.5$. No values are provided for the APEX telescope, since it does not provide an image.}\\
\tablefoottext{b}{In channels of width $4 \sigma_{\text{vel}}$ km s$^{-1}$ and centered at 0 km s$^{-1}$.}
\end{table*}

The study of dusty star-forming galaxies (DSFG) at high redshifts \citep[$z>1$; ][]{casey14} has become a relevant tool to understand the evolution and growth of the most massive galaxies in the universe. These galaxies
play a key role as nurseries of new stars, with star formation rates (SFR) ranging from $\sim 500$ to 1000 M$_\odot$ yr$^{-1}$ and far-infrared luminosities $L_{\text{FIR}}$ $> 10^{12} L_\odot$ \citep{barger12,magnelli12,swinbank14,dacunha15,spilker16,miettinen17,michalowski17}. 
DSFGs are typically found to be at redshifts $z=2-4$ \citep{barger99,ivison02,chapman05,aretxaga07,chapin09,wardlow11,yun12,smolcic12, weiss13,simpson14,chen16,brisbin17}, with a high redshift tail extending out to $z\sim6-7$ \citep{riechers13, weiss13, strandet16, strandet17}; and contribute significantly to the cosmic SFR history \citep[e.g.,][]{magnelli12, sargent14}.
The large SFRs observed in DSFGs yield large stellar masses of $M_*\sim 10^{11} M_\odot$ in $<1$ Gyr \citep{dye08,hainline11,michalowski14,ma15} and can only be sustained by massive molecular gas reservoirs, with typical observed gas masses of $M_{\text{gas}} \sim 3-5 \times 10 ^{10}$ $M_\odot$ \citep{frayer98,greve05,danielson11,bothwell13,sharon13,riechers13,rawle14,aravena16a,huynh17}. The main sources of radiation in these objects are thermal continuum emission from dust grains and line emission from molecular transitions within the interstellar medium (ISM).

Despite their large bolometric luminosities, the detection and study of DSFGs are made difficult not only due to the heavy dust obscuration that usually makes them inconspicuous at optical and near-IR wavelengths, but also due to the weakness of their spectral lines in the redshifted far-infrared regime. 
 
 Here, gravitational lensing provides a solution to both these issues, acting as a natural cosmic telescope to magnify faint galaxies that would otherwise be too difficult to detect. Since gravitational lensing is a geometric effect, sources close in projection to the caustics, are significantly more magnified than sources far from caustics. As a consequence the magnification ($\mu$) of the background galaxy will be the same at all frequencies. However, because the lensed galaxy has a finite size, the magnification will vary across the galaxy \citep{blandford92,hezaveh12, serjeant12}. This is known as differential lensing and can have consequences in the interpretation of the observed integrated properties of lensed galaxies \citep{serjeant12,hezaveh12}. 


The large-scale millimeter survey of the southern sky conducted with the South Pole Telescope \citep[SPT;][]{carlstrom11}, covers 2500 square degrees of the sky, and has unveiled $\sim 100$ of millimeter bright gravitationally lensed DSFGs \citep{vieira10,vieira13}. Follow-up with the Atacama Large Millimeter/submillimeter Array (ALMA) showed that these sources are located at very high redshifts, typically ranging from $z=2-6$ \citep{weiss13,strandet16}. All of these sources are associated with intense starburst activity, and provide an ideal laboratory to study the conditions of galaxy build-up very early in the Universe \citep{spilker15, gullberg15,bethermin16,bothwell16,ma15,ma16,strandet17,marrone18,litke18}.

\begin{figure*}[tbh!]
\includegraphics[width=1\textwidth]{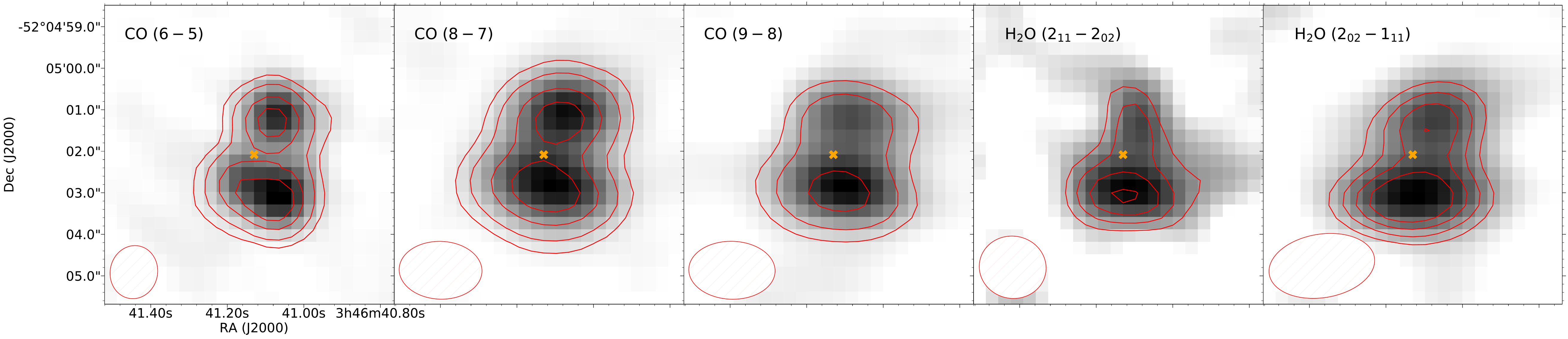}
\caption{Maps of the line transitions detected in SPT0346-52 obtained with ALMA observations, integrated over $4\sigma_{\text{velocity}}$ km s$^{-1}$ channels. For H$_2$O $2_{11}-2_{02}$ and H$_2$O $2_{02}-1_{11}$ red contours are the $2,4,6,8\sigma$ intervals. For CO($6-5$), CO($8-7$) and CO($9-8$) red contours are the $5,10,20,30\sigma$ intervals. The synthesized beam is shown in the bottom-left corner as a red ellipse. The orange cross shows the center of the lensing galaxy. 
\label{fig:map}}         
\end{figure*}

Line emission from the rotational transitions of the carbon monoxide (CO) molecule has been widely used to provide information about the ISM, particularly at high redshift \citep{solomon05, carilli&walter13}. The CO line emission has been observed to trace the distribution and amount of molecular hydrogen in local star forming regions and nearby galaxies \citep{omont07}. Its multiple transitions trace a wide range of physical environments within the cold ISM, and therefore constitutes a useful tool to study the physical properties of the molecular gas in galaxies through modeling of their CO spectral line energy distribution \citep[CO SLED; e.g.,][]{weiss07}. The CO excitation level depends on temperature and gas density, with mid-$J$ to high-$J$ CO lines tracing denser and warmer gas than low-$J$ CO lines \citep{rosenberg15}. A complementary tracer of dense molecular gas is the H$_2$O line emission. H$_2$O transitions trace the dense, warm gas where the radiation produced by young stars or by Active Galactic Nuclei (AGN) increases the temperature of the dust. Since most of the H$_2$O lines fall at rest-frame far-infrared frequencies, ground-based observations of these lines in high redshift galaxies have flourished in recent years, providing an excellent complement to CO studies \citep{omont13,yang16}. 

CO and H$_2$O observations of high-redshift galaxies have increased dramatically in the last few years. However, most studies are not able to accurately measure physical parameters from CO SLED modeling in individual galaxies since this requires multiple CO transitions that trace the full range of excitation conditions. Even among studies that have many CO lines \citep[e.g.,][]{weiss07,riechers13}, spatially resolved observations of many transitions remain scarce \citep[e.g.][]{bothwell13,spilker14}.

The millimeter-bright, gravitationally lensed galaxy SPT0346-52 at $\text{z}=5.6556$ is among the most distant and intrinsically luminous DSFGs known to date. This system seems to be forming stars close to the Eddington limit for a starburst, with a SFR surface density of $\sum_{\text{SFR}}=1540\pm130$ M$_\odot$ yr$^{-1}$ kpc$^{-2}$ using the IR-SFR relation described in \citet{kennicutt98} \citep{hezaveh13,spilker15, ma16}. After correcting for lensing effects its far-IR luminosity yields $L_{\text{FIR}}=(3.6\pm0.3) \times 10^{13} L_\odot$, and thereby is one of the most luminous sources in the SPT DSFG sample \citep{spilker15, ma15, ma16}. Despite the highly concentrated star formation activity, {\it Chandra} observatory X-ray observations provide no evidence indicating the presence of an AGN activity \citep{ma16}. Based on measurement of its CO $(J=2-1)$ line emission, SPT0346-52 was found to have a large molecular gas reservoir of $\sim 8.2 \times 10^{10} \text{M}_\odot$, necessary in sustaining its large IR luminosities \citep{spilker15, aravena16a}.

To understand the gas and dust distribution of this source, \citet{spilker15} used observations of the 870$\mu$m dust continuum emission (rest-frame 130$\mu$m) and the CO($2-1$) line emission with angular resolution of $\sim0.4"$ from ALMA and the Australia Telescope Compact Array (ATCA), respectively. After modeling the effects of the gravitational lensing, these data yield an effective source-plane resolution of $\sim$1 kpc at the redshift of SPT0346-52. Their observations suggest that the dust has a significantly more compact distribution than the diffuse cold gas traced by the low-$J$ CO emission. However, the modest sensitivity of the low-$J$ CO observations did not allow them to study the velocity structure of the gas and thus the actual nature of this source. To understand the relationship between the molecular gas and the star formation process it is necessary to directly trace the denser phases of the molecular gas, resolving the molecular gas both spatially and kinematically. 

\cite{litke18} used a pixellated lensing reconstruction tool \citep{hezaveh16} in SPT0346-52 of $\sim$0\farcs15 ALMA [CII] ($\lambda$ =157.74 $\mu$m) observations, where two spatially and kinematically separated components were found to be connected by a bridge of gas with a size of $\sim$1.6 kpc, which is consistent with a galaxy merger. \cite{dong18} observed and reconstructed the emission of CO(8$-$7) in this source with ALMA $\sim$0\farcs4 data, obtaining an intrinsic CO circularized size of $0.73$ kpc, apparently more compact than [CII].

In this work, we present the observation of CO($10-9$) besides sensitive $J_{\text{up}}=6-9$ CO and H$_2$O($2_{02}-1_{11}$ and $2_{11}-2_{02}$) ALMA imaging of SPT0346-52 at a source-plane resolution of $\sim1$ kpc and with a high SNR. These observations spatially and kinematically resolve both the bright $^{12}$CO and the H$_2$O lines, as well as the cold dust continuum at the observed wavelengths of 2 mm and 3 mm ($300$ and $450\mu$m rest-frame, respectively). Along with the previous resolved observations of CO(2$-$1) and dust continuum at 870$\mu$m, our measurements allow us to (i) reconstruct the molecular gas emission in the source plane, (ii) derive their emitting sizes, resolve the gas distribution along the velocity axis and compare with previous observations, (iii) compare the spatial distributions of the cold and dense molecular gas and dust, and (iv) explore the physical conditions of the ISM in the source-plane through radiative transfer modeling. Hereafter, we adopt a WMAP9 $\Lambda$CDM cosmology with $\Omega_m=0.286$, $\Omega_{\Lambda}=0.713$ and $H_0=69.3 \textrm{ km s}^{-1} \textrm{ Mpc}^{-1}$ \citep{hinshaw13}.

\section{Observations}

\begin{figure*}[h!]
    \centering
    \begin{subfigure}
        \centering
        \includegraphics[width=0.3\linewidth]{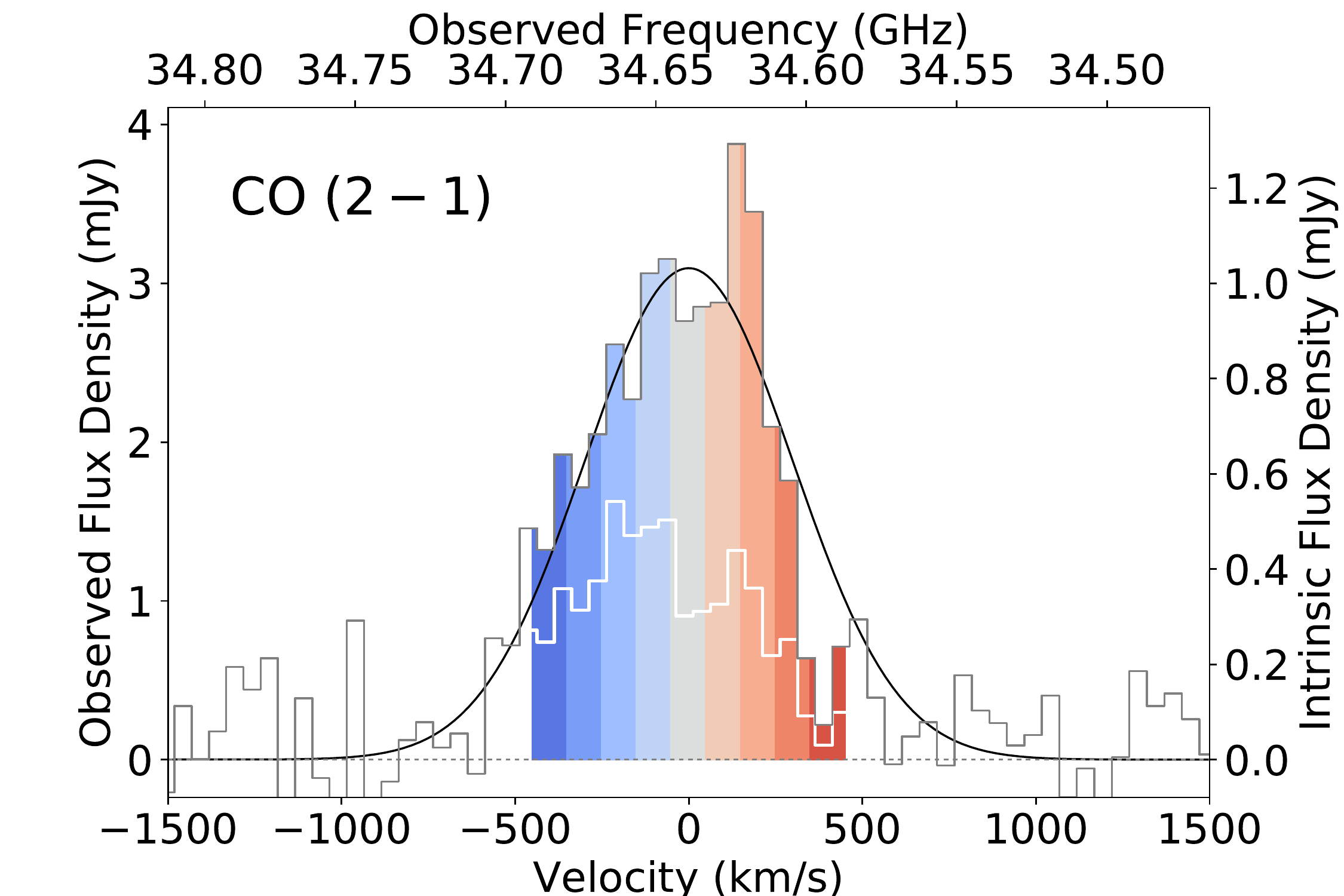} 
    \end{subfigure}
    \hfill
    \begin{subfigure}
        \centering
        \includegraphics[width=0.3\linewidth]{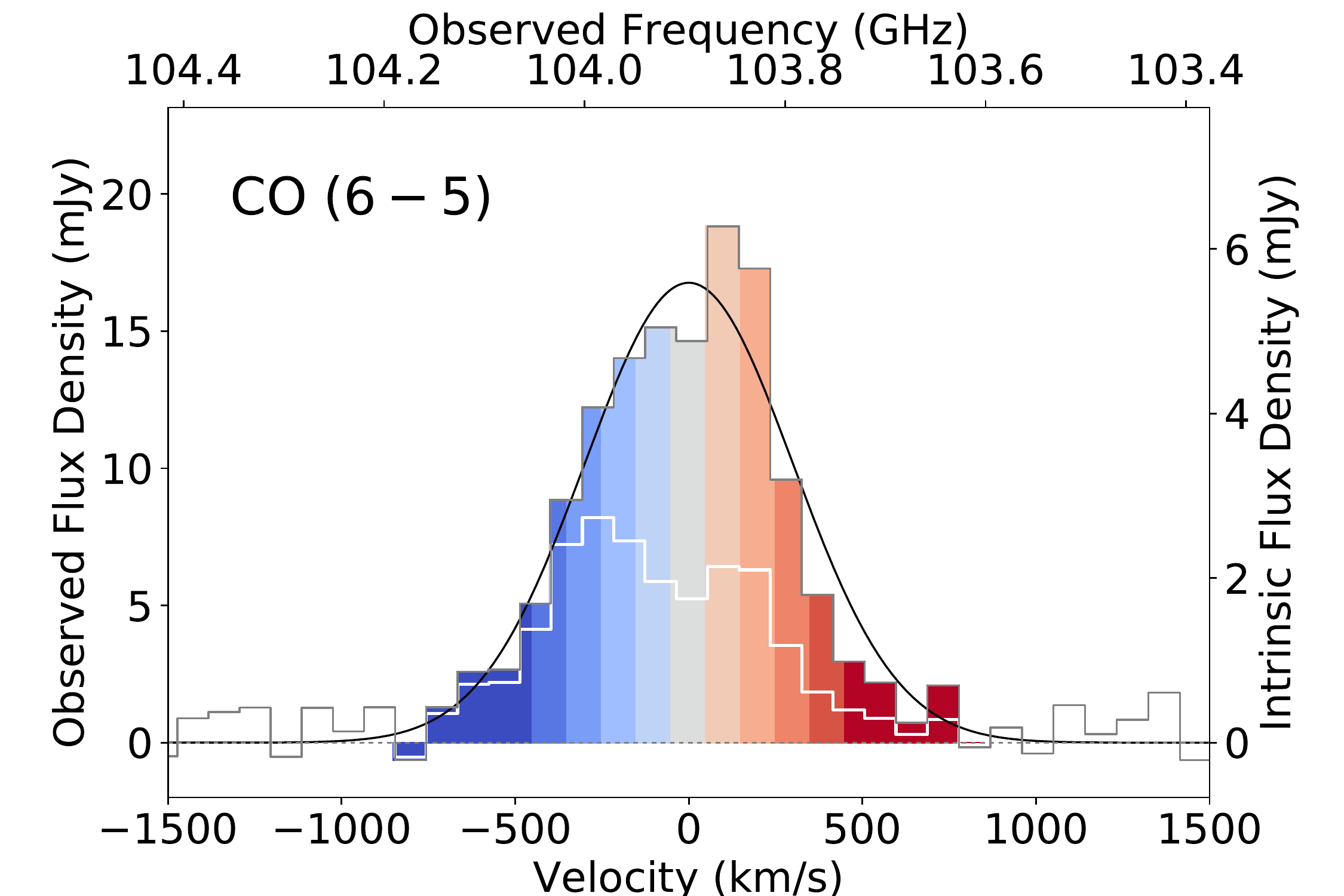} 
    \end{subfigure}
    \hfill
    \begin{subfigure}
        \centering
        \includegraphics[width=0.3\linewidth]{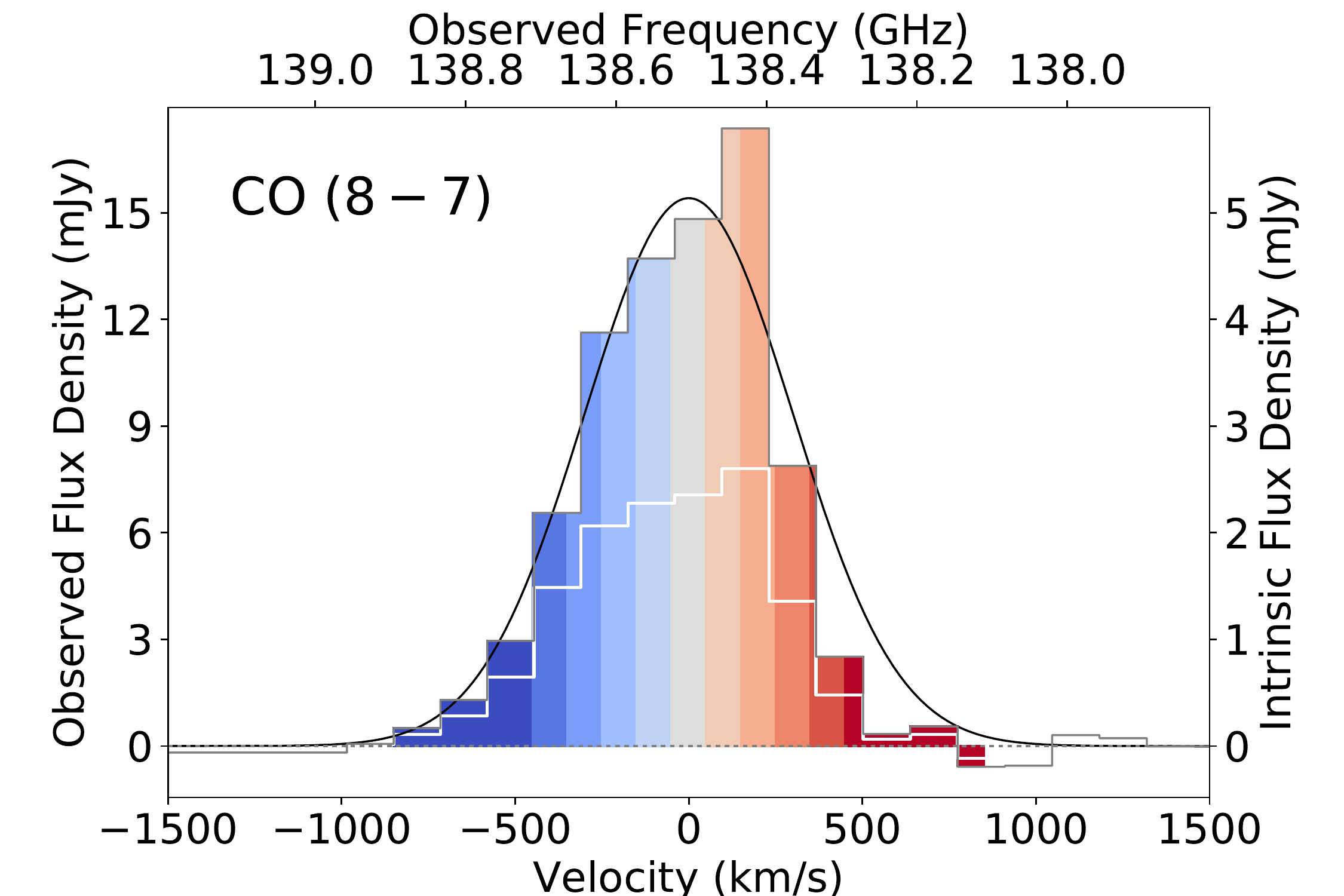} 
    \end{subfigure}
    \vspace{1cm}
    \begin{subfigure}
        \centering
        \includegraphics[width=0.3\linewidth]{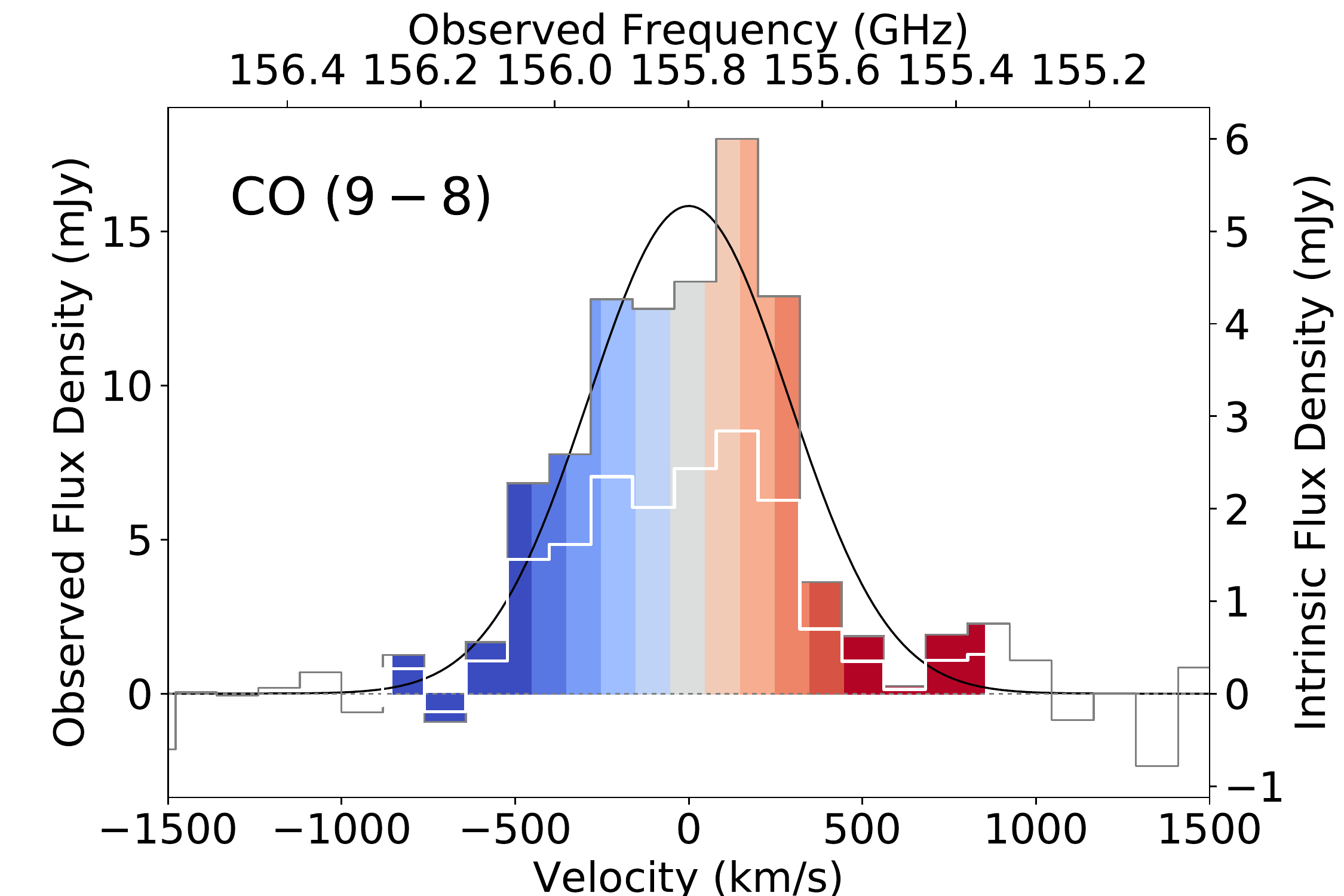} 
    \end{subfigure}
    \hfill
    \begin{subfigure}
        \centering
        \includegraphics[width=0.3\linewidth]{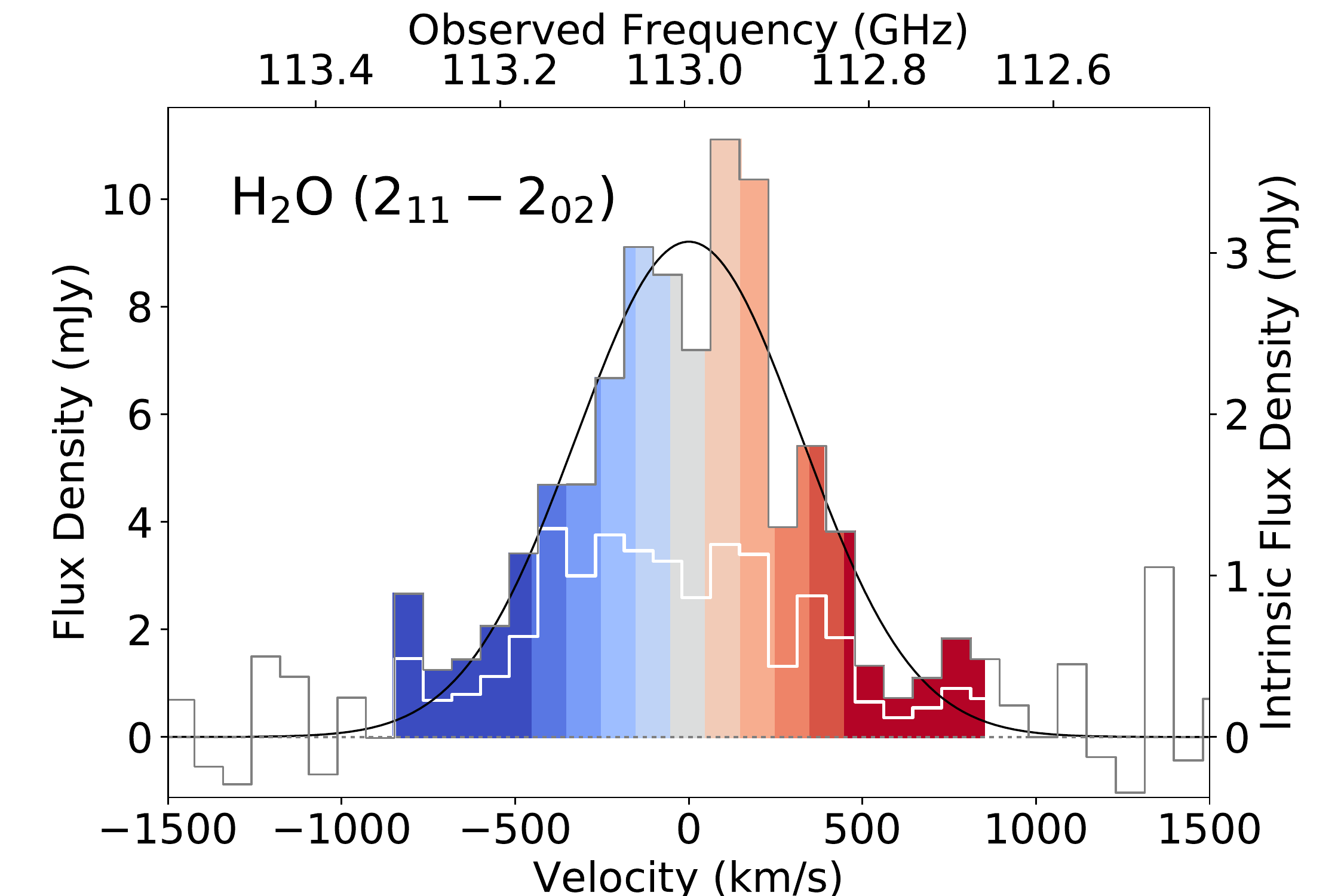} 
    \end{subfigure}
    \hfill
    \begin{subfigure}
        \centering
        \includegraphics[width=0.3\linewidth]{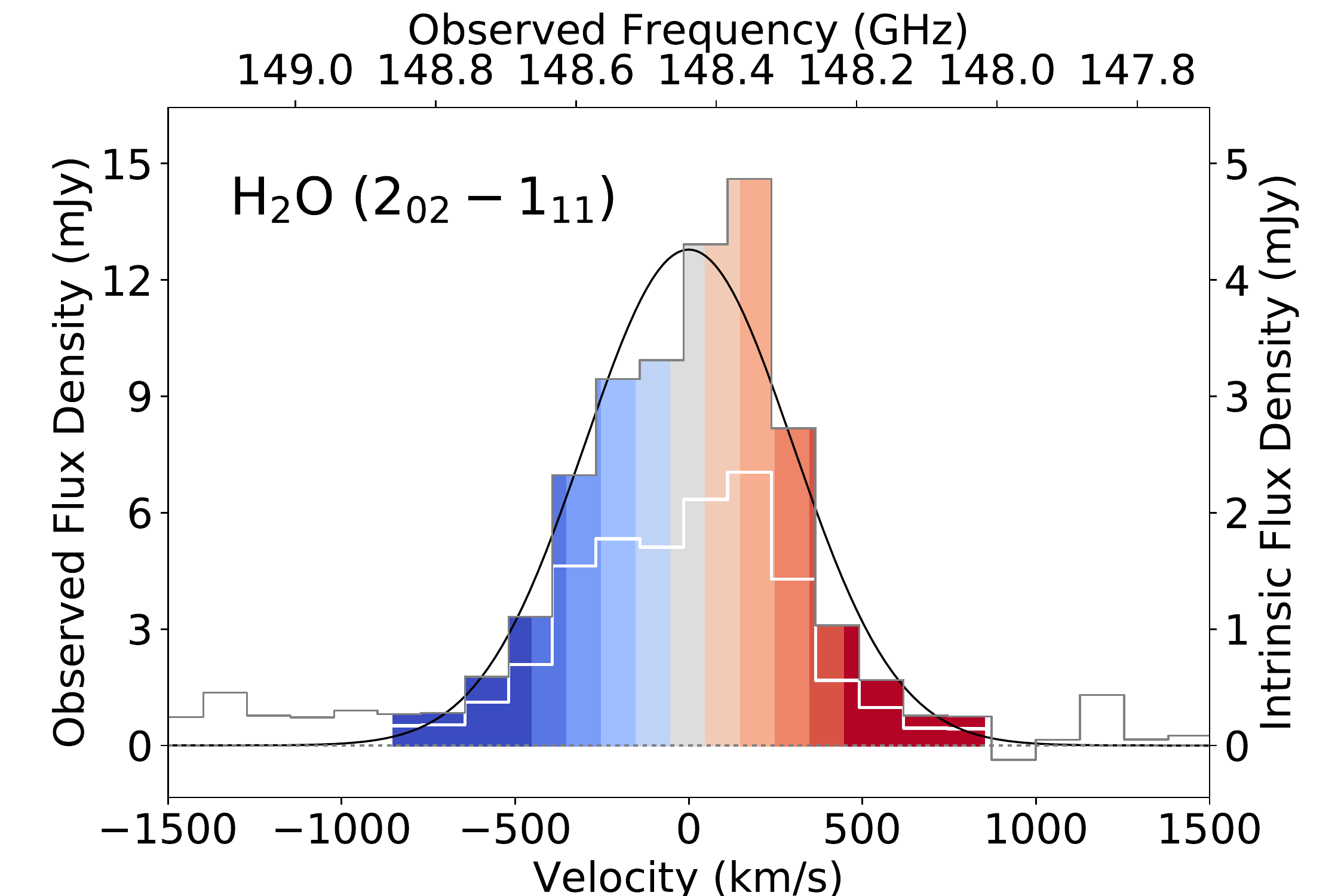} 
    \end{subfigure}    
\caption{Spectra of the line transitions detected in SPT0346-52 obtained with the ATCA and ALMA observations. Single Gaussian fits are represented by black lines, centered at the observed frequencies. The white line shows the intrinsic flux using the values of magnification described in Table \ref{tab:sourcemodelH2O} and \ref{tab:sourcemodelCO}. The color coding denotes the different channels used in the lens modeling. Each channel is 100 km s$^{-1}$ wide, except for the edge channels which are made wider, 400 km s$^{-1}$, to improve on their low SNR.}
\label{fig:spectrum}
\end{figure*}

In this section, we describe details of the molecular line observations of SPT0346-52 at $z=5.7$.

\subsection{ALMA Observations}
Observations of SPT0346-52 were performed with ALMA bands 3 and 4 as a part of projects 2013.1.00722.S and 2015.1.00117.S during ALMA cycles 2 and 3, respectively (PI: M. Aravena). Cycle 2 and 3 observations were conducted during April and May of 2015, and during April of 2016, respectively. 

 The observations were configured to cover the redshifted emission from faint molecular gas lines in the rest-frame far-IR SED \citep[e.g. including HCN, HCO$^+$, $^{13}$CO; see][]{spilker14}. However, in this study, we focus on the brighter, higher significance molecular lines detected of $^{12}$CO and H$_2$O, which allow us to perform detailed lensing source reconstruction. Analysis of the fainter lines will be published elsewhere. A summary of the ALMA observations in this study is provided in Table \ref{tab:summary}.

For each observing band (3 and 4), three frequency tunings were used. Each frequency tuning was observed using four spectral windows (SPWs) in Frequency Division Mode (FDM) and a bandwidth of 1.875 GHz per SPW. We used online spectral averaging that led to a spectral resolutions of 7.8125 MHz and 15.6250 MHz per channel for the band 3 and 4 frequency setups, respectively. These correspond to velocity resolutions of 23 and 31 km s$^{-1}$ per channel at observing frequencies of 100 and 150 GHz, respectively.

Following standard procedures, bright quasars and Mars were used for bandpass, flux and phase calibration, including J0411-5149, J0334-401 and J0519-4546.

The data were reduced and imaged using the Common Astronomy Software Application package \texttt{CASA} \citep{mcmullin07}. We performed an independent reduction and calibration of the data, however we did not see differences with respect to the ALMA pipeline calibration. Time ranges and antennas showing bad visibilities were flagged accordingly, following standard procedures. We created data cubes at different channel resolutions, averaging by 2, 4 and 8 original channels. The visibilities were deconvolved using the \texttt{CLEAN} algorithm. Since our target is detected significantly (SNR $>20$), we masked this source using a tight circular aperture around it with a radius of $5''$, cleaning down to $2\sigma$, where $\sigma$ corresponds to the rms noise level of each channel of the data cube. We tested various weighting schemes, and finally used the Briggs weighting with a robustness factor of 0.5, to have a good trade-off between resolution and sensitivity. The resulting maps are shown in Figure \ref{fig:map}. We note that most of the results of this paper do not depend on the actual weighting scheme used or image resolution obtained, since most of our lens modeling was performed on the visibility plane. 

We performed aperture photometry using an aperture diameter of $7''$ centered at the central lensing galaxy position, using the cubes averaged by 4 channels. We subtracted the continuum emission from the line profiles by fitting a power law to the extracted spectra.

\subsection{APEX Observations}

We used the Atacama Pathfinder Experiment (APEX) single-dish telescope under Max-Planck Institute time (Programme ID: 099.F-9525B) and its SEPIA band 5 receiver to observe the redshifted emission of the $^{12}$CO $J=10-9$ in SPT0346-52. Observations of this line were motivated by the bright detection of CO(9$-$8) obtained with ALMA. The APEX observations were tuned to a frequency of $\sim 173.077$ GHz. SEPIA band 5 is a dual polarization sideband separating receiver, covering 4 GHz bandwidth for both sidebands. The observations were done in wobbler switching mode, with switching frequency of 1.5 Hz and a wobbler throw of 60 arcsec in azimuth, with a total observing time (including integration and overheads) of 34.1 hrs. The precipitable water vapor during the observations varied between 0.4 and 4.5 mm yielding system temperatures between 140 and 1050 K. Pointing was checked frequently and was found to be stable to within $2\arcsec.5 $. Calibration was done every ~10 min using the standard hot/cold-load absorber measurements. 
The beam size and antenna gain at the observing frequency are $31\arcsec$ and 38.4 Jy K $^{-1}$, respectively \citep{belitsky18}. Offline data calibration and flagging was performed using the \texttt{CLASS} software package, including baseline subtraction of polynomials of order 1 for individual scans and spectral averaging. Calibration flux uncertainties are conservatively estimated to be in the range $15-20\%$. The data were smoothed to velocity resolutions of $\sim60$ km s$^{-1}$.

\subsection{ATCA Observations}

We use previous ATCA observations of the $^{12}$CO $J=$2$-$1 emission line toward SPT0346-52 obtained and described in detail by \citet{spilker15}. In short, these observations were centered at a frequency of 34.64 GHz, and obtained in three different array configurations (providing good {\it uv} coverage over baselines from 100m to 6km) integrating on source for 44.8 hours. This led to an angular resolution of $0.45''\times0.65''$ down to $1\sigma$ sensitivities of 54 $\mu$Jy beam$^{-1}$ per 200 km s$^{-1}$ channels.

\begin{figure} [h!]
\centering
\label{fig:apex}
\includegraphics[scale=0.4]{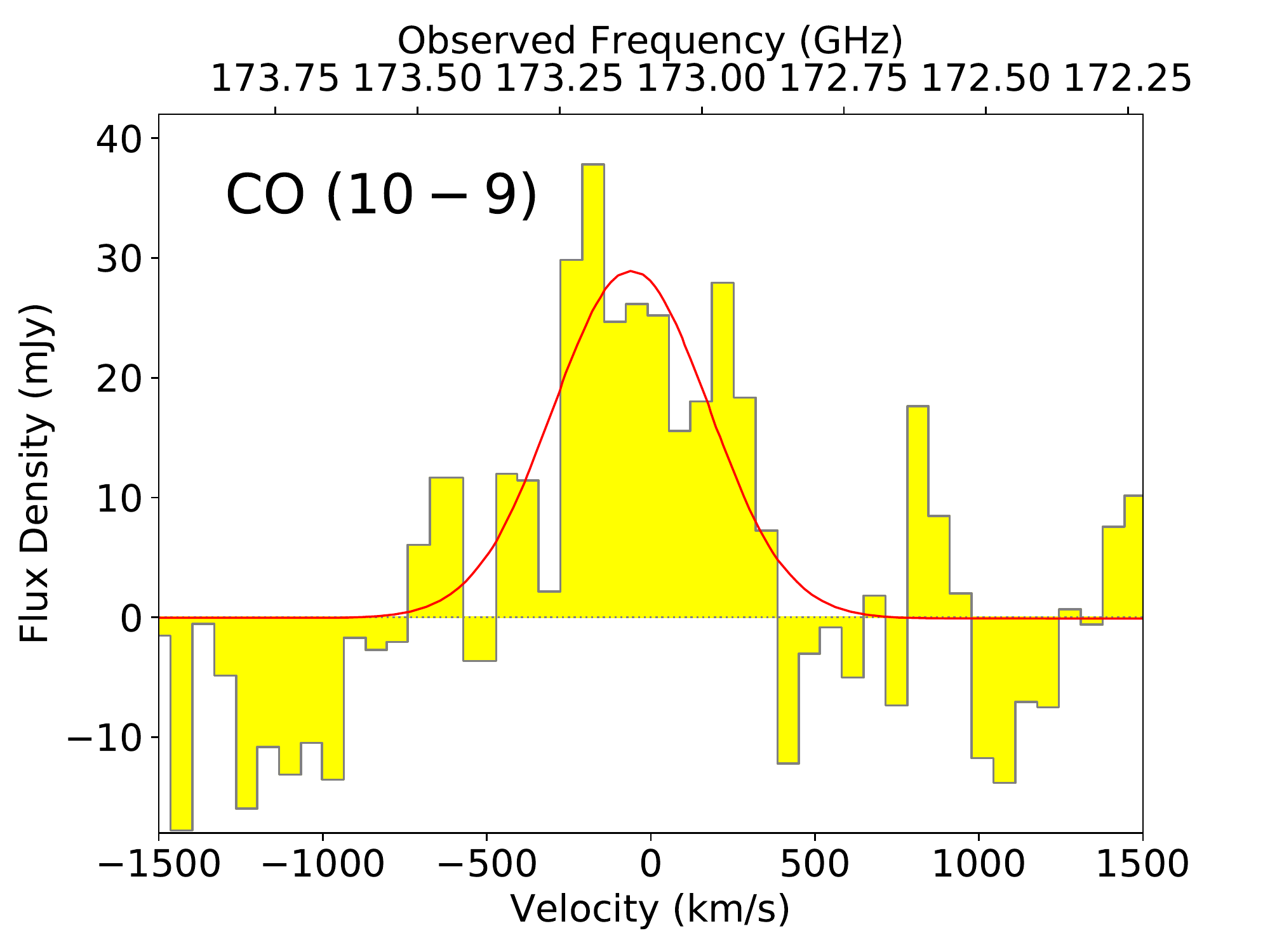}
\caption{Spectrum of the CO($10-9$) line emission observed with APEX of SPT0346-52. The red solid curve shows a Gaussian fit to the line profile. The strength of this line is about 2 times larger than that of CO(9$-$8), suggesting contamination by the a neighboring emission line.}
\end{figure}

\section{Results}

\begin{figure*}[h!]
\includegraphics[width=1\textwidth]{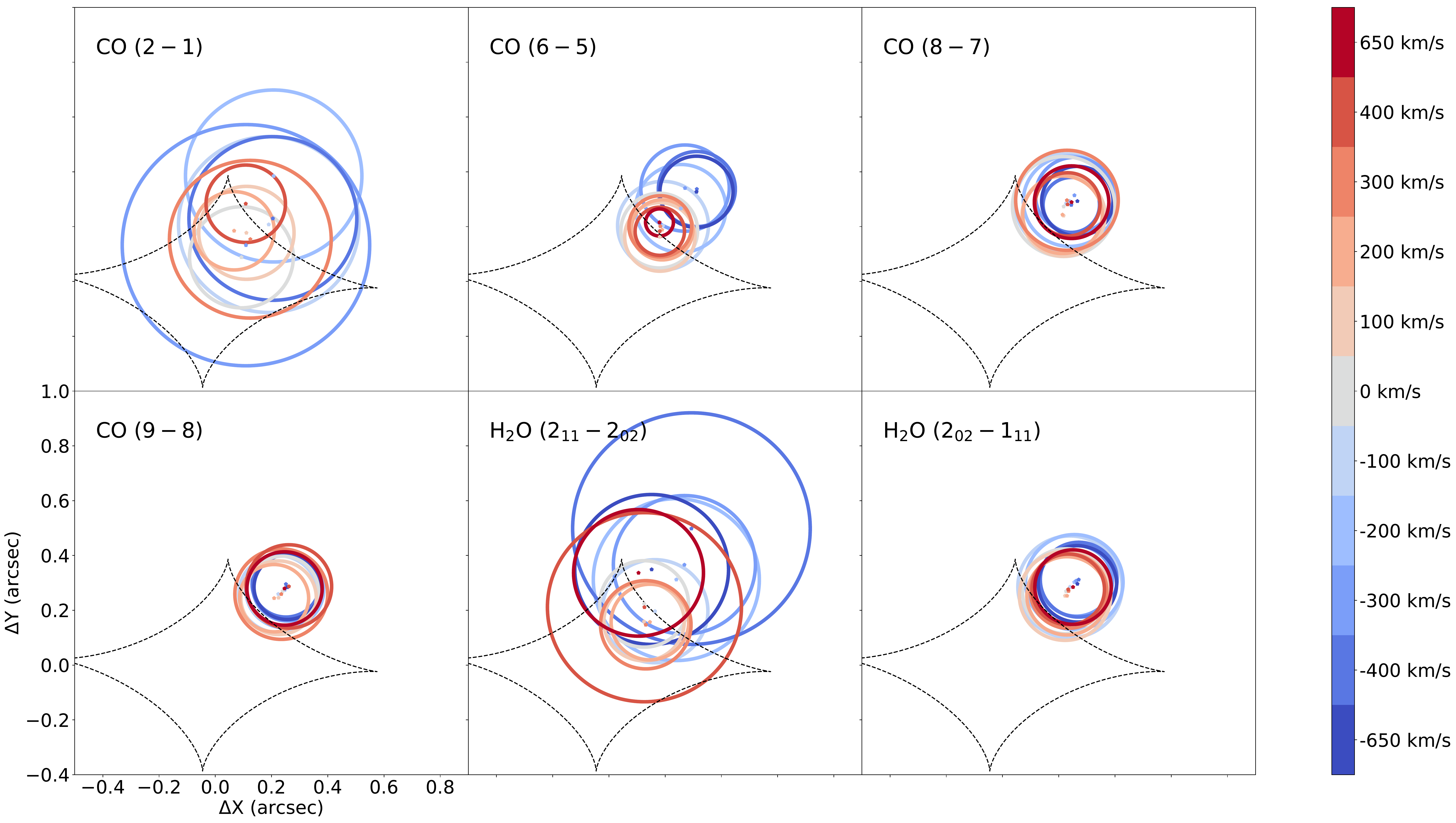}
\caption{Reconstruction of the source plane in each velocity bin as described in the text. The velocity bins are color coded, following the colorbar to the right, and as also represented in Fig \ref{fig:spectrum}. The circles represent the size of the reconstructed emission in each velocity bin, defined by the half light radius as defined in the text. The color coded central dots, show the reconstructed source position at the respective velocity bin. The dotted black line shows the lensing caustic curve from the lens model of \citet{spilker16}.
\label{fig:sourceplane}}
\end{figure*}

\begin{table*}[h!]
\caption{Observed Parameters}              
\label{tab:observedparam}      
\centering                                      
\begin{tabular}{cccc}          
\hline\hline             
{Line} & {$I_{\text{obs,line}}$} & {FWHM} & {$L'_{\text{obs,line}}$} \\
{(kpc)} & {(Jy km s$^{-1}$)} & {(km s$^{-1}$)} & {($10^{10}$ K km s$^{-1}$ pc$^{2}$)}\\
\hline
CO(2$-$1) 					& 2.15$\pm$0.15	& 613$\pm$30	& $60.3\pm4.2$ \\
CO(6$-$5) 					& 11.56$\pm$0.53	& 666$\pm$23	& $36.0\pm1.7$ \\
CO(8$-$7) 				& 10.30$\pm$0.31	& 629$\pm$22	& $18.6\pm0.5$ \\
CO(9$-$8) 						 & 11.10$\pm$0.96	& 685$\pm$53	& $15.4\pm1.3$ \\
H$_2$O(2$_{11}-$2$_{02}$)	 & 7.02$\pm$0.45	& 775$\pm$51	& $18.5\pm1.2$ \\
H$_2$O(2$_{02}-$1$_{11}$) 	 & 8.94$\pm$0.39	& 682$\pm$30	& $13.6\pm0.6$ \\
\hline
Continuum & $S_\text{{obs,dust}}$ & & \\
 & (mJy) & \\
\hline
2 mm							& $8.80  \pm 1.35 $& $ - $ & $ - $ \\
3 mm							& $3.06  \pm 0.05 $& $ - $ & $ - $ \\
\hline
\end{tabular}
\end{table*}

All the targeted $^{12}$CO and H$_2$O lines and the dust continuum emission are significantly detected and spatially resolved. Figure \ref{fig:map} shows the integrated line maps, summed up over $1.7\times$ the line full width at half-maximum (FWHM). As expected for a lensed source, each line map shows two lobes which are multiple images of the same source and respectively its line emission. Spectra of each individual emission line are shown in Figure \ref{fig:spectrum}.

The line FWHM and central frequency are computed from single, one-dimensional Gaussian fits to the emission line profiles. Fluxes are computed by directly integrating the continuum subtracted line profiles over the relevant velocity range. The continuum emission is measured using a $7''$ diameter aperture around the source. Table \ref{tab:observedparam} lists the observed measured line and continuum fluxes.

All the line profiles are consistent in width and central frequency within the uncertainties. However, in the case of  CO($10-9$), the single Gaussian fit yields an integrated flux of $\sim 17.3 \pm 2.9$ Jy km s$^{-1}$, which is approximately 2 times brighter than that of CO(9$-$8). The line width of $560 \pm 100$ km s$^{-1}$ is in agreement with the other CO lines. \cite{gonzalezalfonso10} describes a blending between H$_2$O $3_{12} - 2_{21}$ ($\nu_{\text{obs}}$=173.25 GHz) and CO(10$-$9) ($\nu_{\text{obs}}$=173.08 GHz) since both lines lie within $\approx$300 km s$^{-1}$ from each other. From the H$_2$O line ratios observed in NGC253, NGC4945 and Arp220 \citep{liu17}, we estimate a possible contribution in the range $5.6 - 11.5$ Jy km s$^{-1}$ from H$_2$O $3_{12} - 2_{21}$ to the detected feature, which would imply a flux density  of $5.8 - 11.7$ Jy km s$^{-1}$ for CO(10$-$9). However, if this H$_2$O line were blended with CO(10-9) we would expect some broadening of the detected line feature, which we do not see. This, however, can have been missed due to the moderate significance of the detection. We note that the expected calibration uncertainties (of the order $15-20\%$) cannot account for the excess flux.

\subsection{Lensed Source Reconstruction}

To understand the intrinsic geometry of the background high redshift galaxy in SPT0346-52, we have performed source plane reconstructions through a strong lens model \citep[\texttt{VISILENS};][]{spilker14}. The method relies on modeling the visibilities instead of images and simultaneously determining self-calibration phases \citep{hezaveh13}. The emission regions have been assumed to have a circularly symmetric Gaussian light profile with four free parameters: flux density F$_S$, scale radius R$_S$, and position offsets from the lens center X$_S$ and Y$_S$. Since we are interested in the intrinsic sizes of emission regions, the choice of a Gaussian profile compared to e.g. S\'ersic or other commonly used source profiles is justified in order to minimize the number of free source parameters.

The lensing mass distribution used was fixed to that obtained by \cite{spilker16} using ALMA 870$\mu$m observations with a resolution of $\sim 0.5''$. It is comprised of a Singular Isothermal Ellipsoid (SIE) plus an external shear component. The fixed parameters that define the lens model are its Einstein radius $\theta_{\text{E,L}}$, SIE position angle $\phi_{\text{L}}$, ellipticity $e_{\text{L}}$, external shear strength $\gamma$ and position angle $\theta_\gamma$ (see Table \ref{tab:lenstable}). The central position of the lens (a nuisance parameter) was fixed to that obtained through a lens model that used the highest signal to noise line while keeping the other parameter fixed to those of \citet{spilker16}. Each of the lens-modeled lines in SPT0346-52 are split into nine velocity bins with a width of 100 km s$^{-1}$, ranging from $-400$ to $400$ km s$^{-1}$. The lensed emission from each one of these bins has been de-projected into the source plane. Due to the lower SNR, in the case of the wings of the lines, the first and last velocity bin widths have been increased by a factor of four (from $-850$ to $-450$ km s$^{-1}$ and from $450$ to $850$ km s $^{-1}$, respectively).

The derived source plane arrangement is shown in Figure \ref{fig:sourceplane}. The radius of each circle is the half-light radius defined as $r_{1/2}=\sqrt{-2 \ln{0.5}}\sigma$, where $\sigma$ is the Gaussian dispersion of the source light profile. The rest of the derived parameters are shown in Table \ref{tab:sourcemodelH2O} and \ref{tab:sourcemodelCO}. 
Finally, we also performed a reconstruction of each line using a single, wide velocity channel (from $- 2\sigma_{vel}$ to  $2\sigma_{vel}$) as a benchmark for comparison. The obtained source parameters are shown in Table \ref{tab:sourcemodel}. 

\begin{table}[h!]
\caption{Fixed lens model parameters from \citet{spilker16}\label{tab:lenstable}}
\centering
\begin{tabular}{c|lc}
\hline \hline
Parameter & Value \\ 
\hline
$\theta_{\text{E,L}} $ 			& $0.\arcsec979 \pm 0\arcsec.007$ \\
$e_{\text{L}}$ 					& $0.52 \pm 0.03$ \\
$\phi_{\text{L}}$ 				& $71^\circ \pm 1^\circ$  \\
$\gamma$						& $0.12 \pm 0.01$  \\
$\phi_\gamma$ 					& $122^\circ \pm 3^\circ$ \\
\hline
\end{tabular}
\end{table}

\begin{table}[h!]
\caption{Reconstruction Parameters \label{tab:sourcemodel}}
\centering
\begin{tabular}{ccc}
\hline \hline
{Line} & {$\mu_{\text{line}}$} & {$r_{\text{eff}}$} \\
{} & {} & {(kpc)} \\
 \hline
CO(2$-$1) 						& 8.30 $\pm$ 0.39 & 1.15 $\pm$ 0.08  \\
CO(6$-$5) 						& 6.09 $\pm$ 0.07 & 0.92 $\pm$ 0.02  \\
CO(8$-$7) 						& 4.86 $\pm$ 0.24 & 0.86 $\pm$ 0.05  \\
CO(9$-$8) 						& 5.36 $\pm$ 0.03 & 0.79 $\pm$ 0.02  \\
H$_2$O $2_{11}-2_{02}$	& 6.78 $\pm$ 0.94 & 1.30 $\pm$ 0.27  \\
H$_2$O $2_{02}-1_{11}$ 	& 5.28 $\pm$ 0.03 & 0.95 $\pm$ 0.02  \\
\hline
Continuum & & \\
\hline
$2$mm & $5.04 \pm 0.09$ &   $0.73 \pm 0.03$\\
$3$mm & $4.63 \pm 0.03$ &   $0.79 \pm 0.02$\\
\hline
\end{tabular}
\end{table}

\subsection{Intrinsic CO line emission regions}

\begin{figure*}[h!]
\includegraphics[width=1\textwidth]{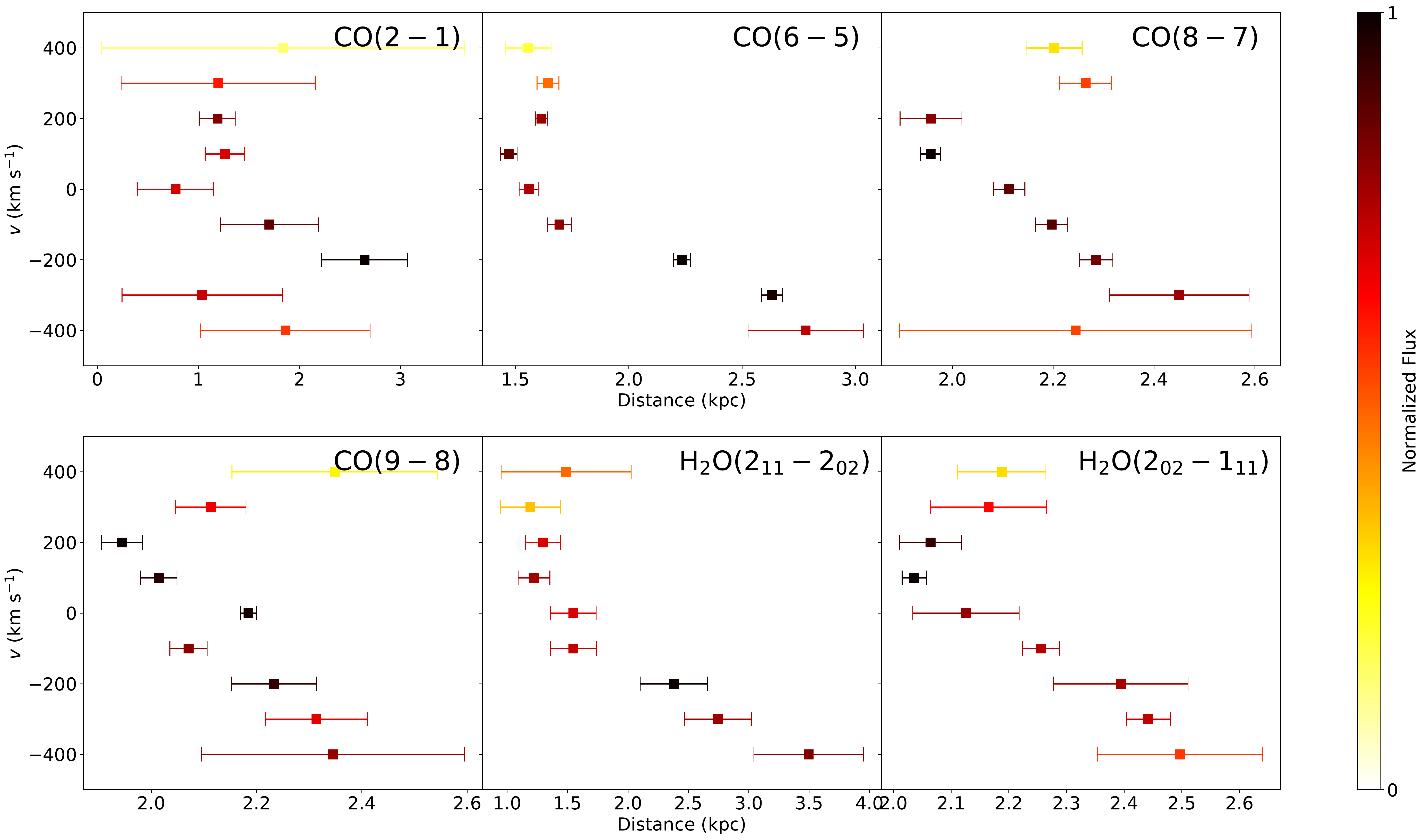}
\caption{Position relative to the center of the lens, for each velocity bin. The color bar shows the normalized flux density of each transition line.}
\label{fig:velvsposition}
\end{figure*}

\begin{figure}
\includegraphics[width=0.45\textwidth]{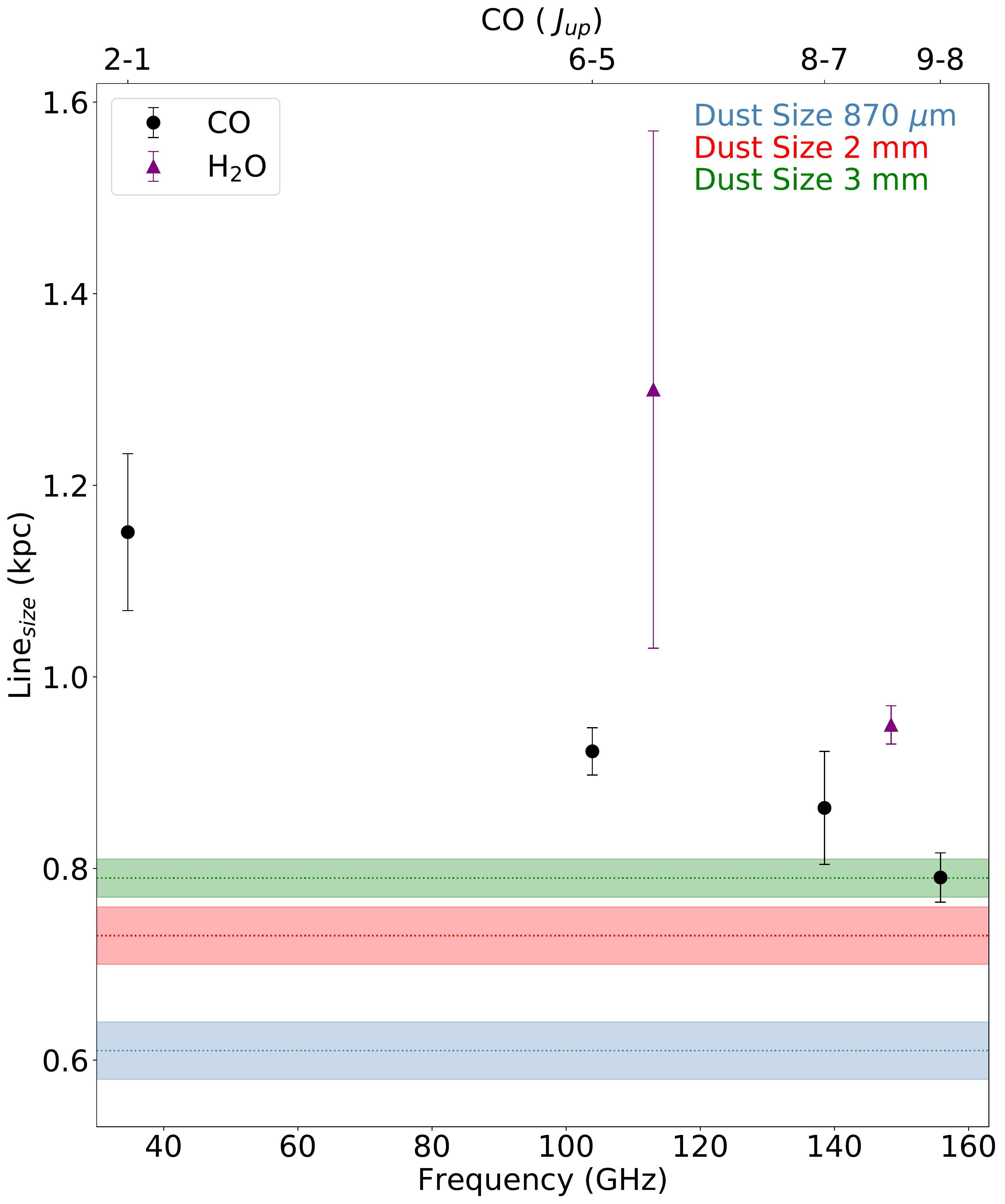}
\caption{Dependency of the reconstructed source size (half light radius) of the line emission regions with transition/frequency. Black dots shows reconstructed CO line emission regions as a function of the $J$ transition and the observed frequency. Purple triangles shows reconstructed H$_2$O line emission as a function of the observed frequency. The dotted lines represent the size (r$_{1/2}$) of the dust for different wavelengths, blue for 870  $\mu$m, red for 2mm and green for 3mm continuum emission.}
\label{fig:COsizes}
\end{figure}

As seen in Figure \ref{fig:sourceplane} and more evidently in Figure \ref{fig:COsizes}, the size of the CO line emitting regions are more compact for higher rotational transitions, with CO(2$-$1) coming from a region of $r_{1/2}=1.15$ kpc down to CO(9$-$8) coming from a region of $r_{1/2}=0.79$ kpc. Even though this is expected and has been observed in other sources such as M82, where the low-$J$ CO is dominated by the diffuse outer regions while transitions $J_{\text{up}}$>5 are emitted from the central regions \citep{weiss05a}, it could not have been observed without the enhancement to the resolution produced by strong lensing in this system, allowing also to have a picture not only of the different fluxes as well the variation of the CO sizes through the quantum number. 

We observe some differences on the structure between different CO transitions with respect to both reconstructed size and position of each velocity bin. In CO(2$-$1) a trend of increasing emission region size is seen with decreasing radial velocity (see Table \ref{tab:sourcemodelCO} and Fig.  \ref{fig:sourceplane}): red wings of the lines appear to emit from regions three times more compact than those emitting the blue wings, with ranges $0.9 \lesssim r_{1/2} \lesssim 2.7$ kpc. This trend is also hinted in  CO(6$-$5), however, only in the highest velocity bins. Within the uncertainties, however, all CO ($J_{\text{up}} > 2$) are consistent with no size variation per velocity bin. We note that this variation seen only in CO(2$-$1) may be due to the fact that this line has been observed with ATCA at a significantly lower significance than the rest of the lines. 

Regarding relative position of the reconstructed emission regions per velocity bins, we see in Fig.  \ref{fig:velvsposition} that CO(2$-$1) does not follow any specific trend. On the other hand, in all of the three higher CO transitions studied, a clear gradient is observed suggesting a rotating disk-like structure for these lines. The projected extension of this line of sight structure of the emission regions is largest for CO(6$-$5) ($\sim$1.2 kpc) and smallest for CO(9$-$8) ($\sim$0.4 kpc), consistent with the integrated lines reconstruction sizes (Fig. \ref{fig:COsizes}).

\subsection{H2O lines}

In the H$_2$O transitions, we also observe some differences in structure. As shown in Table \ref{tab:sourcemodelCO} and Fig.  \ref{fig:sourceplane} the H$_2$O $(2_{11}-2_{02})$ transition shows variation of reconstructed sizes per velocity bin, with no clear correlation (albeit with high uncertainty in reconstructed sizes) thus suggesting a merger scenario. On the other hand, much like the CO ($J_{\text{up}} > 2$) lines reconstruction, the rotational H$_2$O $(2_{02}-1_{11})$ transition line reconstruction shows no size variation per velocity bin within the uncertainties.

As seen in Fig. \ref{fig:velvsposition} both H$_2$O lines show line of sight structure with projected extensions of $\sim$2.3 kpc for H$_2$O $(2_{11}-2_{02})$ and $\sim$0.5 kpc for H$_2$O $(2_{02}-1_{11})$. Note that the lower SNR of H$_2$O $(2_{11}-2_{02})$ line coupled with the significantly larger reconstructed sizes of the regions implies less robust reconstructed center measurements and thus the projected extension.

\subsection{Size, magnification and bias}

Gravitational lensing is a geometrical effect: sources close in projection to the ``infinite magnification'' caustics, are significantly more magnified than sources far from caustics and compact sources are magnified more than extended sources. As a direct consequence, the magnification and distortion of sources with non-isotropic morphologies has a strong correlation not only with their shape, but their projected positions relative to the lensing potential. These two (sometimes competing) effects are shown in Tables \ref{tab:sourcemodelH2O} and \ref{tab:sourcemodelCO} where the most compact regions or the regions closest to the caustics are not necessarily the most magnified. As such, this supports the idea that the intrinsic structure of lensed sources detected in sub-millimeter surveys may play a significant role in their detection \citep[e.g.][]{hezaveh12}.

\section{Discussion}

\subsection{CO Ratios}

\begin{figure*}
\includegraphics[width=1\textwidth]{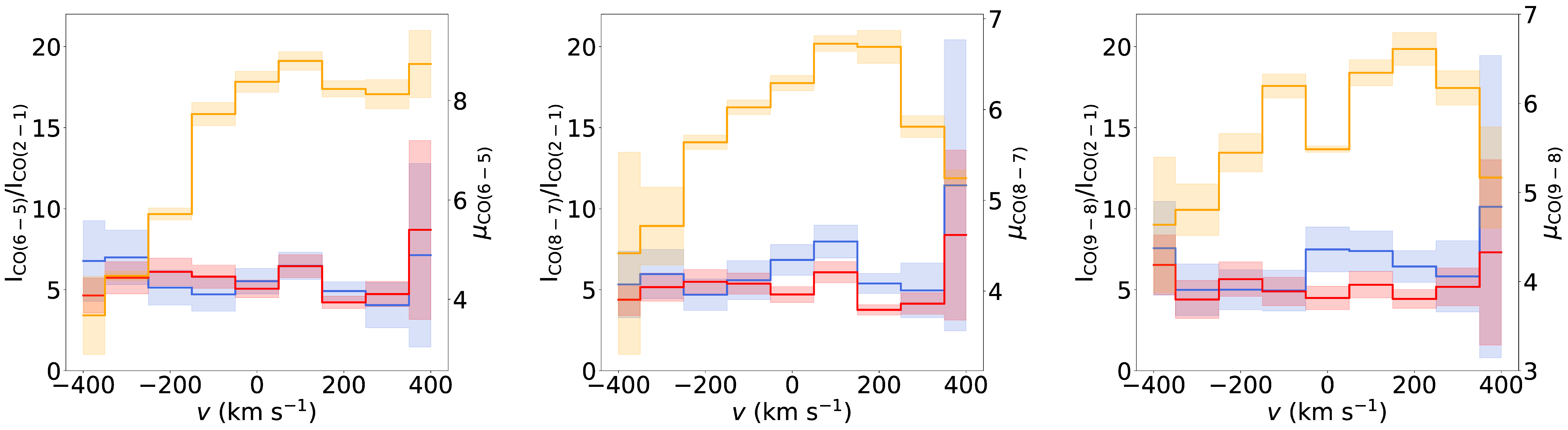}
\caption{CO ratio of intrinsic (blue line) and observed emission line intensities} (red line) across the line profile for CO($6-5$)/CO($2-1$) (left panel),  CO($8-7$)/CO($2-1$) (middle panel) and  CO($9-8$)/CO($2-1$) (right panel). Orange line represent the magnification factor across the line profile for CO($6-5$), CO($8-7$) (middle panel) and CO($9-8$) (right panel).
\label{fig:gasratio}
\end{figure*}

Since our model are not able to capture the morphology of each line, and thus we cannot investigate the effects of differential magnification in the projected plane, we explore the variation of the magnification and CO line ratios as a function of velocity (i.e. along the line of sight). 

Figure \ref{fig:gasratio} shows the variation with velocity of the magnification and the ratio of each CO line with respect to CO($2-1$). Also shown is the observed line luminosity for reference. We find that the magnification factor obtained for each CO line varies throughout the line profile (Figure \ref{fig:gasratio}), showing a consistent behavior among all three lines. The magnifications seem to increase from $-400$ km s$^{-1}$ to a peak at around $100-200$ km s $^{-1}$, decreasing at larger velocities.

Figure \ref{fig:gasratio} also shows the CO ratios ($I_{\rm CO J\geq6}/I_{\rm CO(2-1)}$) of the observed and intrinsic line fluxes, as a function of velocity. The magnifications vary significantly across velocities (a factor of 2 for CO($6-5$)), however the CO line ratios present milder differences for both observed and intrinsic measurements. While the observed line ratios show little variation with velocity, the intrinsic line ratios show non-negligible gradients along the line of sight. This behaviour is broadly consistent with the results presented by \cite{dong18}, where fairly constant CO line ratios for this source across velocities are found, even though they use partially different datasets. Note, however, that \cite{dong18} study brightness temperature ratios with coarser velocity resolution, whereas here we show the intensity ratios (their Fig. 9). Neglecting the $+400$ km s$^{-1}$ velocity bin, which has large uncertainties, there is a tendency of having a larger line ratios at $v\sim0-200$ km s$^{-1}$, suggesting higher excitation in this velocity range. Conversely, for $v<0$ we observe a roughly constant ratio with variation less than $25 \%$ for $I_{8-7}/I_{2-1}$ and less than $17\%$ for $I_{9-8}/I_{2-1}$.

 It is interesting to note that, neglecting the H$_2$O($2_{11}-2_{02}$) line which has the higher uncertainties, the H$_2$O line has a size compatible within the uncertainties with those of the CO($6-5$) or CO($8-7$) lines, and significantly higher than CO($9-8$). This contrasts the results from low-z ULIRGs, where the H$_2$O emission arises from particularly dense, dust-obscured regions \citep{gonzalezalfonso10, koenig17}. One possible explanation for this could be that the high-$J$ CO and H$_2$O lines, which in general trace similarly dense star forming regions, are essentially probing different physical mechanisms within the galactic environment. High-$J$ CO lines trace collisionally excited molecular gas, whereas the H$_2$O lines trace gas that is illuminated by IR photons (radiative IR pumping). Therefore, if the IR radiation extends beyond the collisionally excited regions (typically traced by CO), the H$_2$O molecules located farther away could be excited, leading to larger sizes. Mapping of other H$_2$O transitions at high(-er) angular resolution will be able to solve this puzzle.

\subsection{CO SLED}

\begin{figure}[tbh!]
\includegraphics[width=0.45\textwidth, trim={1cm 0 10cm 0}]{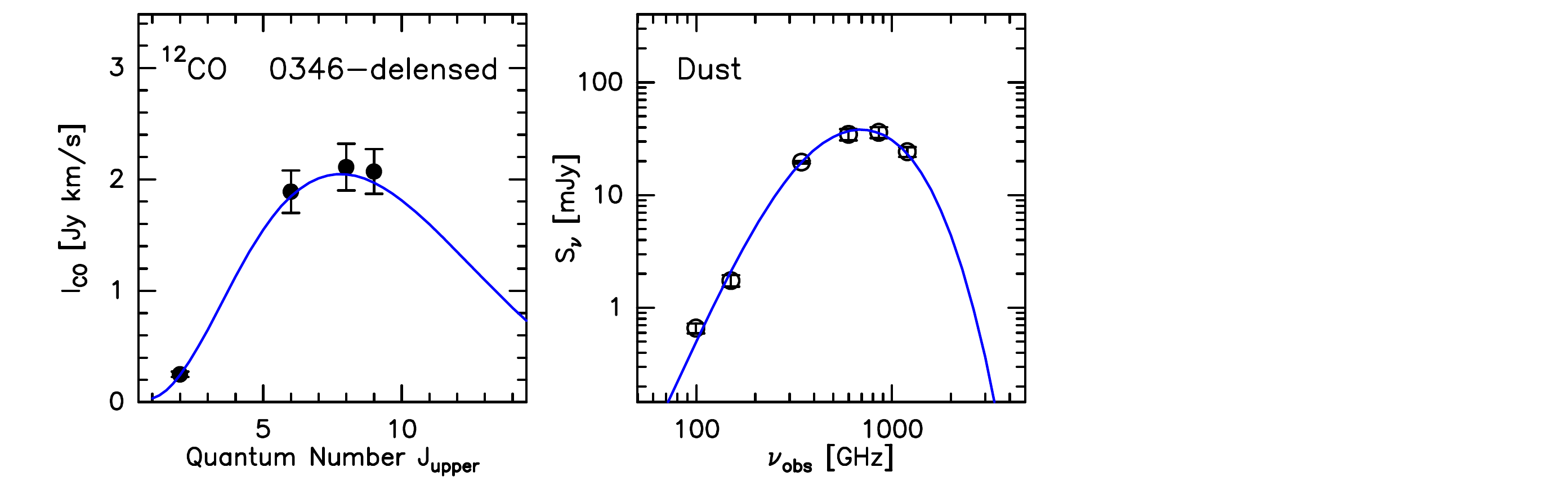}
\caption{Result of the fitting of the CO-SLED (left panel) and SED (right panel). The filled circles are the CO rotational lines and the open circles represent the continuum fluxes obtained in this work and from \citet{spilker16} and \citet{ma16}. The blue line in both panels shows the best-fit model.}
\label{fig:LVGmodel}
\end{figure}

We study the physical conditions of the ISM in SPT0346-52 by modeling the CO lines and continuum emission, using the method described in \citet{weiss07}. This method is based on a spherical, multi-component, large velocity gradient (LVG) model. Here the dust continuum and the CO rotational transition line emission are linked via the gas column density  $N_{\text{H$_{2}$}}$ which is defined by
\begin{equation}
N_{\text{H$_{2}$}}=3.086 \times 10^{8} n(\text{H$_{2}$}) \frac{\Delta V_{\text{turb}}}{dv/dr}
\end{equation}
where $\Delta V_{\text{turb}}$ is the turbulence linewidth, $n(\text{H$_{2}$})$ the gas density and $dv/dr$ the velocity gradient of the LVG model. From this, we can obtain the total gas mass from the LVG model, using
\begin{equation}
M_{\text{LVG}}=0.21 r_0^2 n(\text{H$_{2}$}) \frac{\Delta V_{\text{turb}}}{dv/dr} (M_\odot),
\label{eq:gasmass}
\end{equation}
where $r_0^2$ is the equivalent source radius which depends on angular size distance to the source ($D_{\text{A}}$) and the magnified solid angle $(\Omega_{\text{app}})$. 

\begin{table} [ht!]
\centering
\caption{Summary of the radiative transfer model\label{tab:LVGresult}}
\begin{tabular}{c|lc}
\hline  \hline 
{Parameter} & {Value} \\
\hline 
$log n_{\text{H$_{2}$}}$ 		& $4.9    \pm  1.6$ log(cm$^{-3}$)\\
$T_{\text{kin}}$ 			& $43	\pm 14$ K \\
$T_{\text{dust}}$ 			& $29.1 	\pm 0.8 $ K \\
$\kappa_{\text{vir}}$ 		& $3.4 	\pm 2.3$ \\
$M_{\text{H$_2$}}$ 			& $3.9  	\pm 2.2 \times 10^{11} M_\odot $ \\
$L_{\text{FIR}}$ 			&$2.2 		\pm 0.2 \times 10^{13} L_\odot $ \\
\hline 
\end{tabular}
\end{table}

We run the model using the intrinsic flux densities of all CO lines and FIR continuum fluxes. We include our 2mm and 3mm ALMA continuum measurements, the ALMA 0.87mm flux \citep{spilker16}, and the 250$\mu$m, 350$\mu$m and 500$\mu$m from Herschel/SPIRE observations \citep{ma16}. To correct the Herschel SPIRE fluxes, we use the magnifications obtained at 0.87mm \citep{spilker16}. Since the CO line widths are relatively uniform, we assume a fixed CO line width, corresponding to the weighted average line width value (642 km s$^{-1}$).

Figure \ref{fig:LVGmodel} shows the CO spectral line energy distribution (SLED) and continuum measurements. The CO SLED shows a peak close to CO(8$-$7), however it appears to stay roughly constant toward higher $J$ transitions, indicating a highly excited gas component which could extends beyond $J=10-9$. The ISM properties derived from the best fit radiative transfer model are summarized in Table \ref{tab:LVGresult}. 

The kinetic temperature ($T_{\text{kin}} \sim 43$ K) is larger than the dust temperature ($T_{\text{dust}} \sim 29$ K). This could suggest that the ISM is being affected by an external mechanical energy input. This result is in agreement with the virial velocity gradient ($\kappa_{\text{vir}} \sim 3.4$) obtained from the model, where the uncertainties are affected by the cloud geometry and the density profile. We note that most parameters in this model, including the CO abundance and intrinsic source sizes, are left as free parameters. Our model yields an intrinsic source size of $4.4\pm2.4$ kpc, a CO chemical abundance of $8\times10^{-5}$, $dv/dr=220$ and $\Delta_{\rm turb}=40$.

Using Equation \ref{eq:gasmass}, we derive a total intrinsic gas mass of $\sim 3.9\pm2.2 \times 10^{11} \text{M}_\odot$, which, despite the large uncertainties, is significantly higher than the molecular gas mass derived using the low-$J$ CO line alone \citet{aravena16a}, and even higher than the dynamical mass estimate of $1.5\times10^{11}\ M_\sun$ obtained from the high resolution CO($2-1$) imaging \citep{spilker15}. This difference of the LVG-based molecular gas mass and the dynamical mass is $10\%$ above the formal $1\sigma$ uncertainties of the LVG measurement. The latter are mostly driven by our inability to trace the peak of the CO SLED with our current observations and thus our radiative transfer model is not fully constrained. 
Note that this discrepancy could also reflect uncertainties in the CO luminosity to gas mass conversion factor used to transform the low-$J$ CO luminosities.

\begin{figure}[tbh!]
\includegraphics[width=0.48\textwidth]{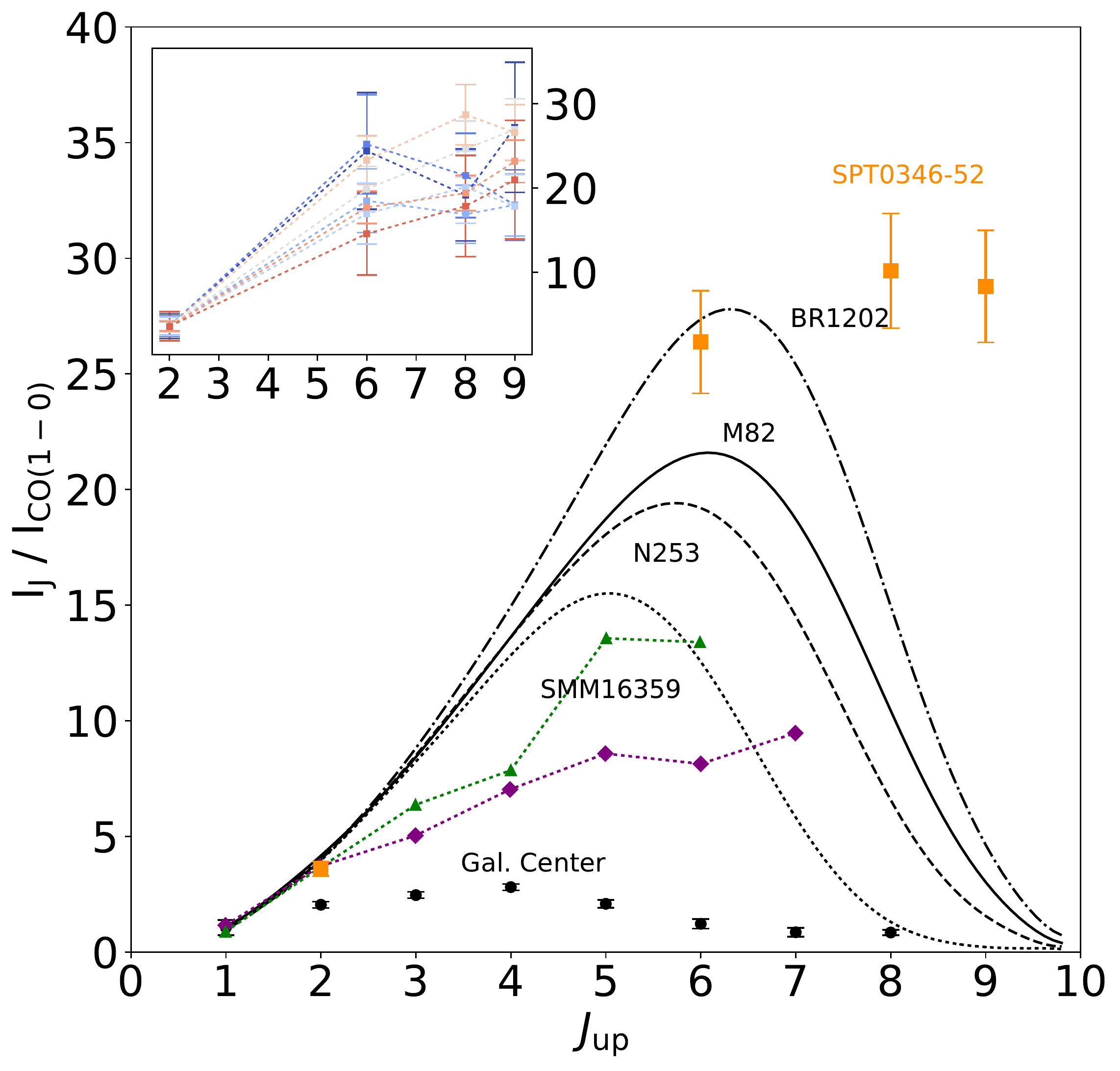}
\caption{Comparison of the CO SLEDs normalized to the CO$(1-0)$ flux density transition with different galaxies. Orange squares represent the obtained CO SLED of SPT0356-52 taking the line flux described in Table \ref{tab:sourcemodel}. Black lines shows BR$1202-0725$ ($z=4.69$, \citep[]{carilli02a,riechers06}), M82 Center \citep{weiss05a},  NGC253 Center \citep{gusten06}, SMM16359 ($z=2.5$, \citet{weiss05b}). The solid circles shows the Galactic Center \citep{fixsen99}, diamonds shows average SLED of DSFGs \citep{bothwell13} and triangles shows SLED of SPT DSFGs \citep{spilker14}.
The inset shows the CO SLED normalized to the CO$(1-0)$ for different velocities in SPT0356-52.}
\label{fig:COratioJ}
\end{figure}

Figure \ref{fig:COratioJ} compares the integrated CO SLED normalized to CO($1-0$) of SPT0346-52 with the literature. For this comparison, we normalize the CO SLED of SPT0346-52 to the CO($1-0$) line intensity by assuming a brightness temperature line ratio $R_{21}=T^B_{21}/T^B_{10}=0.9$ \citep{carilli&walter13}, and using the measured CO($2-1$) line intensity to obtain the CO($1-0$) intensity from $I_{10}=I_{21} (\nu_{21}/\nu_{10})^2$ $R_{21}^{-1}$.

SPT0346-52 has clearly a higher CO excitation than local  starburst galaxies, and even luminous quasars. Interestingly, SPT0346-52 seems to be an extreme case when  compared to most distant DSFGs \citep[e.g.,][]{bothwell13, spilker14, canameras18}, with a CO SLED similar to that of HLFS-3 at $z=6.3$ \citep{riechers13}. This is exemplified by the large gas density obtained through the radiative transfer model. Confirmation of this higher excitation nature will require observations of higher J CO transitions. The inset of Figure \ref{fig:COratioJ} shows what could be interpreted as a dependency of the excitation with velocity, however, as shown by the uncertainties, it is consistent with no variation.

\subsection{H$_2$O line emission}

Recent systematic observations of the H$_2$O emission lines showed that these lines are among the most luminous in the far-infrared spectra of galaxies \citep[e.g.][]{omont13, yang13, yang16}. Furthermore, the H$_2$O line luminosities were found to correlate tightly with their total IR luminosities. This has been found to be an almost direct consequence of IR pumping \citep{vanderwerf11, yang13, yang16}, and thus the H$_2$O emission likely traces the far-infrared radiation field generated in dense star forming regions in galaxies. While some galaxies that host AGN show lower $L_{\mathrm{H}2\mathrm{O}}/L_\mathrm{IR}$ ratios, no real dependence has been found on the existence of AGN or lack thereof \citep{yang13}. 

\begin{figure}[tbh!]
\includegraphics[width=0.44\textwidth]{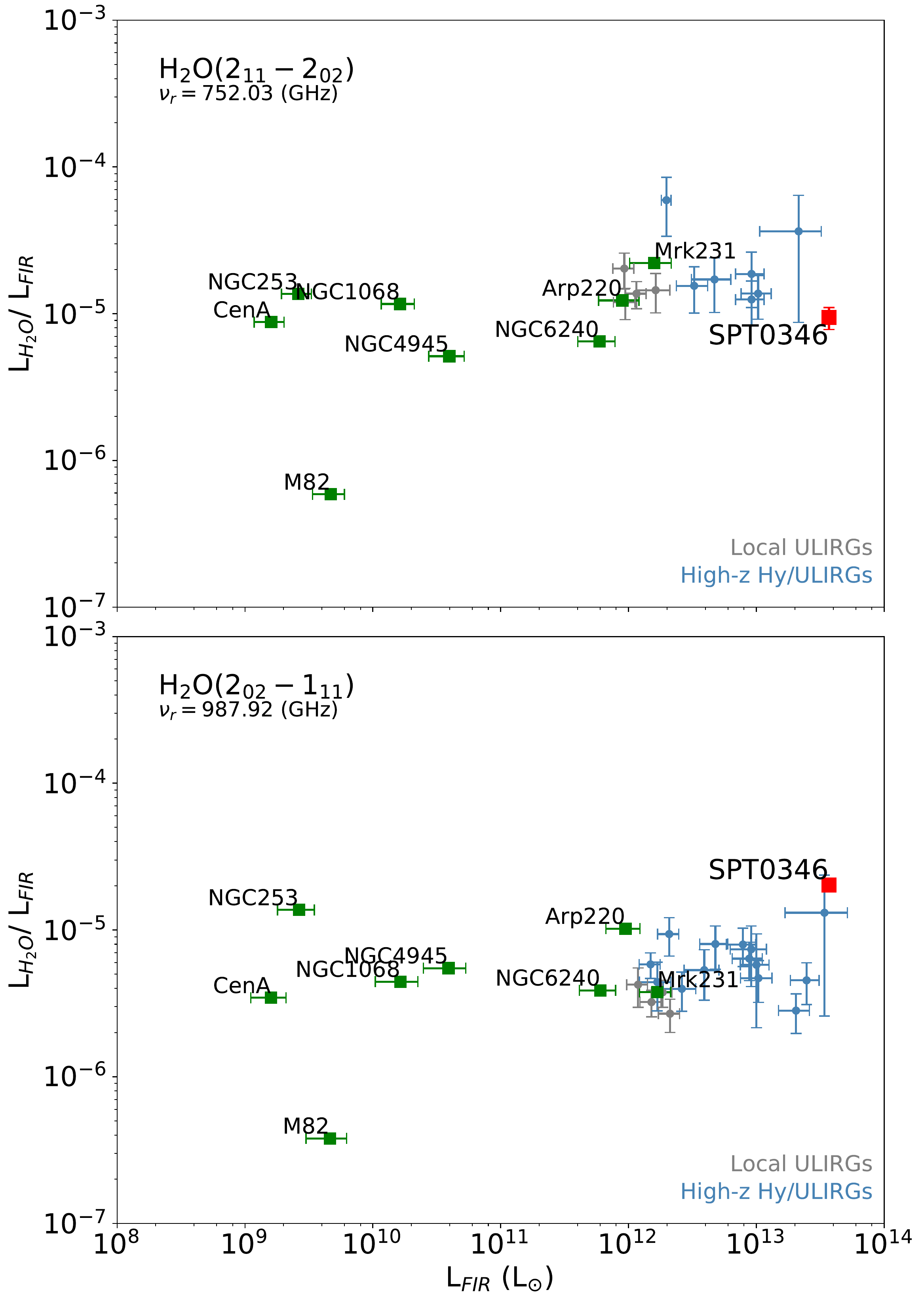}
\caption{Correlation between the H$_2$O/FIR and FIR luminosities for a sample of galaxies in the literature, compared to SPT0346-52. Green squares represent local starburst galaxies observed with {\it Herschel} \citep{liu17}, grey circles show local ULIRGS and blue circles show high-redshift Hy/ULIRGs \citep{yang16}. The large red square symbol shows the location of SPT0346-52. The uncertainties in some of these measurements are smaller than the size of each data point. SPT0346-52 follows the trend, suggesting the presence of radiative pumping of the H$_2$O molecules.}
\label{fig:LfirvsLH2O}
\end{figure}

As previously shown, we significantly detected two H$_2$O emission lines toward SPT0346-52. This is one of the most distant objects in which H2O has been detected \citep[see also:][]{riechers13}. Figure \ref{fig:LfirvsLH2O} shows the H$_2$O and IR luminosities of SPT0346-52 compared to other galaxies at low and high redshift in the literature. We find that our source, closely follows the correlation between these two quantities, suggesting the presence of IR radiative pumping of the H$_2$O molecules in the ISM.  In addition, we find no signs for signature pointing to the presence of AGN based on this correlation, in agreement with previous results from \cite{ma16}. The close match of the spatial location of the reconstructed CO(6$-$5) and H$_2$O lines  in the source-plane (Figure \ref{fig:sourceplane}) and the agreement between the line widths of the H$_2$O and the high-$J$ CO transitions observed, supports the idea that H$_2$O and the high-$J$ CO are co-located, presumably in a region of intense star formation.

\subsection{Gas Tracers}

The reconstruction results presented here are consistent with those presented in \cite{litke18} and \cite{dong18}. The kinematics of our CO(6$-$5) reconstruction is similar to the [CII] reconstruction in \cite{litke18} (see Fig. 4), where the $v > 0$ part of the gas lies directly on the caustic line while for $v < 0$ the emission shows a displacement of $\sim 1$ kpc. Our CO$(J=8-7 \text{ and } 9-8)$ and H$_2$O(2$_{02}-$2$_{11}$) reconstruction indeed show a more compact structure with a separation of $\sim 0.5$ kpc between the bluest and reddest part of the gas, near the ``bridge'' that \cite{litke18} found between the two [CII] lobes. 

Restricting the lines to the range with higher SNR (within $\pm450$ km s$^{-1}$), we  observe that for these three lines the largest part of the flux comes from $v\geq0$, with $55\pm4\%$, $54\pm2\%$ and $57\pm2\%$ of the line emission coming from this range for the CO($9-8$), CO($8-7$) and H2O($2_{02}-1_{11}$), respectively. 

Conversely, the three remaining lines that have an extended structure (CO $J=(2-1)$, (6$-$5) and H$_2$O$(2_{11}-2_{02})$), show a peak of the flux at $\sim -200$ km s$^{-1}$ but with a second increment of the flux between $\sim 100 - 200$  km s$^{-1}$. 
It is important to note that a direct comparison between the reconstructed source shape with this work is not possible since in \cite{dong18} the ellipticity (and position angle) in the source plane reconstruction for each velocity bin in the line have been left free in the fit. Since magnification is defined exactly as image to source area ratios, complete freedom in ellipticity can result in differing fitted sizes and magnifications, making the contrast of recovered sizes per velocity not directly comparable. In fact, magnifications and sizes measured by \citet{dong18} in each velocity bin are consistent, though systematically larger and smaller, respectively, than those presented here. Fitted central positions, on the other hand, are only mildly affected by this freedom and are indeed compatible with those shown here. 

\section{Summary and Conclusions}
We have presented mid to high-$J$ CO and H$_2$O transition lines in SPT0346-52 using ALMA Band 3 and Band 4 observations. Using the \texttt{VISILENS} code and lens model parameters obtained in \cite{spilker16} we reconstructed the lensed galaxy in the source-plane, and have analyzed its intrinsic ISM properties. The main results of this study are the following:

(i) We derived intrinsic sizes, positions and fluxes of multiple CO lines ($J=6-5$, $8-7$ and $9-8$) and H$_2$O ($2_{02}-1_{11}$ and $2_{11}-2_{02}$) in several velocity channels per line. Although there is in general a continuous gradient in the projected extension of the line of sight structure that would suggest a disk-like morphology, at least two emission lines (CO($2-1$) and H$_2$O$(2_{11}-2_{02})$) show significant variations in their reconstructed sizes at resolved velocity bins. Even when these two lines have the largest uncertainties in the reconstructed sizes, it could be consistent with a more complicated structure such as a galaxy merger. This scenario is indeed consistent with that shown by \citet{litke18}. Furthermore, the variation of the normalized flux along the velocity axis in all line transitions (CO and H$_{2}$O) suggests evidence of a disturbed structure, which supports  the scenario of a galaxy merger found by \citet{litke18}.

(ii) The CO sizes show a clear trend of decreasing size with increasing $J$, and the CO sizes for all transitions appear larger than the dust continuum sizes (at rest-frame 450, 300 and 130$\mu$m).

(iii) Our observations suggest differential magnification in the CO lines along the velocity axis. While the observed CO line ratios appear roughly constant with velocity, we find that the intrinsic CO line ratios (CO($6-5$)/CO($2-1$), CO($8-7$)/CO($2-1$) and CO($9-8$)/CO($2-1$)) show a peak at $\sim0-200$ km s$^{-1}$. This could argue for a more excited region within the galaxy.

(iv) CO SLED shows a peak at $J=8-7$, indicating excited gas with a T$_{\text{kin}}\sim 43$ K and T$_{\text{dust}}\sim 29$ K. The intrinsic CO SLED seems to be more excited than local starburst and luminous quasars.

(v) H$_2$O emission follows the correlation of L$_{\text{FIR}}$ $/$ L$'_{\text{H$_{2}$O}}$ in accordance with IR radiative pumping. The resemblance between CO and H$_2$O of the linewidths and the structure of the source plane show that the emission is likely co-spatial, coming from regions of intense star formation.

\begin{acknowledgements}
      Y.A. acknowledge support of Universidad Andr\'es Bello through a Graduate School Fellowship. Y.A. and T.A. acknowledge support by the Ministry for the Economy, Development, and Tourism's Programa Inicativa Cient\'ifica Milenio through grant IC 12009. This paper makes use of the following ALMA data: ADS/JAO.ALMA 2013.1.00722.S and 2015.1.00117.S. ALMA is a partnership of ESO (representing its member states), NSF (USA) and NINS (Japan), together with NRC (Canada), MOST and ASIAA (Taiwan), and KASI (Republic of Korea), in cooperation with the Republic of Chile. The Joint ALMA Observatory is operated by ESO, AUI/NRAO and NAOJ. The National Radio Astronomy Observatory is a facility of the National Science Foundation operated under cooperative agreement by Associated Universities, Inc. The Australia Telescope Compact Array is part of the Australia Telescope National Facility which is funded by the Australian Government for operation as a National Facility managed by CSIRO. This publication is based on data acquired with the Atacama Pathfinder Experiment (APEX). APEX is a collaboration between the Max-Planck-Institut fur Radioastronomie, the European Southern Observatory, and the Onsala Space Observatory. The SPT is supported by the NSF through grant PLR-1248097, with partial support through PHY-1125897, the Kavli Foundation and the Gordon and Betty Moore Foundation grant GBMF 947. S.J., J.D.V. and  D.P.M. acknowledge support from the US NSF under grants AST-1715213 and AST-1716127. S.J. acknowledge support from the US NSF NRAO under grant SOSPA5-001. J.D.V. acknowledges support from an A. P. Sloan Foundation Fellowship. The Flatiron Institute is supported by the Simons Foundation. D.N. was supported in part by NSF Award AST-1715206 and HST Theory Award 15043.0001
\end{acknowledgements}

\bibliographystyle{aa}
\bibliography{aanda.bib}

\newcommand{\noop}[1]{}
\begin{thebibliography}{76}
\expandafter\ifx\csname natexlab\endcsname\relax\def\natexlab#1{#1}\fi

\bibitem[{{Aravena} {et~al.}(2016){Aravena}, {Spilker}, {Bethermin},
  {Bothwell}, {Chapman}, {de Breuck}, {Furstenau}, {G{\'o}nzalez-L{\'o}pez},
  {Greve}, {Litke}, {Ma}, {Malkan}, {Marrone}, {Murphy}, {Stark}, {Strandet},
  {Vieira}, {Weiss}, {Welikala}, {Wong}, \& {Collier}}]{aravena16a}
{Aravena}, M., {Spilker}, J.~S., {Bethermin}, M., {et~al.} 2016, \mnras, 457,
  4406

\bibitem[{{Aretxaga} {et~al.}(2007){Aretxaga}, {Hughes}, {Coppin}, {Mortier},
  {Wagg}, {Dunlop}, {Chapin}, {Eales}, {Gazta{\~n}aga}, {Halpern}, {Ivison},
  {van Kampen}, {Scott}, {Serjeant}, {Smail}, {Babbedge}, {Benson}, {Chapman},
  {Clements}, {Dunne}, {Dye}, {Farrah}, {Jarvis}, {Mann}, {Pope}, {Priddey},
  {Rawlings}, {Seigar}, {Silva}, {Simpson}, \& {Vaccari}}]{aretxaga07}
{Aretxaga}, I., {Hughes}, D.~H., {Coppin}, K., {et~al.} 2007, \mnras, 379, 1571

\bibitem[{{Barger} {et~al.}(1999){Barger}, {Cowie}, {Smail}, {Ivison}, {Blain},
  \& {Kneib}}]{barger99}
{Barger}, A.~J., {Cowie}, L.~L., {Smail}, I., {et~al.} 1999, \aj, 117, 2656

\bibitem[{{Barger} {et~al.}(2012){Barger}, {Wang}, {Cowie}, {Owen}, {Chen}, \&
  {Williams}}]{barger12}
{Barger}, A.~J., {Wang}, W.-H., {Cowie}, L.~L., {et~al.} 2012, \apj, 761, 89

\bibitem[{{Belitsky} {et~al.}(2018){Belitsky}, {Lapkin}, {Fredrixon},
  {Meledin}, {Sundin}, {Billade}, {Ferm}, {Pavolotsky}, {Rashid}, {Strandberg},
  {Desmaris}, {Ermakov}, {Krause}, {Olberg}, {Aghdam}, {Shafiee}, {Bergman},
  {Beck}, {Olofsson}, {Conway}, {Breuck}, {Immer}, {Yagoubov},
  {Montenegro-Montes}, {Torstensson}, {P{\'e}rez-Beaupuits}, {Klein}, {Boland},
  {Baryshev}, {Hesper}, {Barkhof}, {Adema}, {Bekema}, \& {Koops}}]{belitsky18}
{Belitsky}, V., {Lapkin}, I., {Fredrixon}, M., {et~al.} 2018, \aap, 612, A23

\bibitem[{{B{\'e}thermin} {et~al.}(2016){B{\'e}thermin}, {De Breuck},
  {Gullberg}, {Aravena}, {Bothwell}, {Chapman}, {Gonzalez}, {Greve}, {Litke},
  {Ma}, {Malkan}, {Marrone}, {Murphy}, {Spilker}, {Stark}, {Strandet},
  {Vieira}, {Wei{\ss}}, \& {Welikala}}]{bethermin16}
{B{\'e}thermin}, M., {De Breuck}, C., {Gullberg}, B., {et~al.} 2016, \aap, 586,
  L7

\bibitem[{{Blandford} \& {Narayan}(1992)}]{blandford92}
{Blandford}, R.~D. \& {Narayan}, R. 1992, \araa, 30, 311

\bibitem[{{Bothwell} {et~al.}(2016){Bothwell}, {Maiolino}, {Peng}, {Cicone},
  {Griffith}, \& {Wagg}}]{bothwell16}
{Bothwell}, M.~S., {Maiolino}, R., {Peng}, Y., {et~al.} 2016, \mnras, 455, 1156

\bibitem[{{Bothwell} {et~al.}(2013){Bothwell}, {Smail}, {Chapman}, {Genzel},
  {Ivison}, {Tacconi}, {Alaghband-Zadeh}, {Bertoldi}, {Blain}, {Casey}, {Cox},
  {Greve}, {Lutz}, {Neri}, {Omont}, \& {Swinbank}}]{bothwell13}
{Bothwell}, M.~S., {Smail}, I., {Chapman}, S.~C., {et~al.} 2013, \mnras, 429,
  3047

\bibitem[{{Brisbin} {et~al.}(2017){Brisbin}, {Miettinen}, {Aravena}, {Smol{\v
  c}i{\'c}}, {Delvecchio}, {Jiang}, {Magnelli}, {Albrecht}, {Arancibia},
  {Aussel}, {Baran}, {Bertoldi}, {B{\'e}thermin}, {Capak}, {Casey}, {Civano},
  {Hayward}, {Ilbert}, {Karim}, {Le Fevre}, {Marchesi}, {McCracken},
  {Navarrete}, {Novak}, {Riechers}, {Padilla}, {Salvato}, {Scott},
  {Schinnerer}, {Sheth}, \& {Tasca}}]{brisbin17}
{Brisbin}, D., {Miettinen}, O., {Aravena}, M., {et~al.} 2017, \aap, 608, A15

\bibitem[{{Ca{\~n}ameras} {et~al.}(2018){Ca{\~n}ameras}, {Yang}, {Nesvadba},
  {Beelen}, {Kneissl}, {Koenig}, {Le Floc'h}, {Limousin}, {Malhotra}, {Omont},
  \& {Scott}}]{canameras18}
{Ca{\~n}ameras}, R., {Yang}, C., {Nesvadba}, N.~P.~H., {et~al.} 2018, \aap,
  620, A61

\bibitem[{{Carilli} {et~al.}(2002){Carilli}, {Kohno}, {Kawabe}, {Ohta},
  {Henkel}, {Menten}, {Yun}, {Petric}, \& {Tutui}}]{carilli02a}
{Carilli}, C.~L., {Kohno}, K., {Kawabe}, R., {et~al.} 2002, \aj, 123, 1838

\bibitem[{{Carilli} \& {Walter}(2013)}]{carilli&walter13}
{Carilli}, C.~L. \& {Walter}, F. 2013, \araa, 51, 105

\bibitem[{{Carlstrom} {et~al.}(2011){Carlstrom}, {Ade}, {Aird}, {Benson},
  {Bleem}, {Busetti}, {Chang}, {Chauvin}, {Cho}, {Crawford}, {Crites}, {Dobbs},
  {Halverson}, {Heimsath}, {Holzapfel}, {Hrubes}, {Joy}, {Keisler}, {Lanting},
  {Lee}, {Leitch}, {Leong}, {Lu}, {Lueker}, {Luong-van}, {McMahon}, {Mehl},
  {Meyer}, {Mohr}, {Montroy}, {Padin}, {Plagge}, {Pryke}, {Ruhl}, {Schaffer},
  {Schwan}, {Shirokoff}, {Spieler}, {Staniszewski}, {Stark}, {Tucker},
  {Vanderlinde}, {Vieira}, \& {Williamson}}]{carlstrom11}
{Carlstrom}, J.~E., {Ade}, P.~A.~R., {Aird}, K.~A., {et~al.} 2011, \pasp, 123,
  568

\bibitem[{{Casey} {et~al.}(2014){Casey}, {Narayanan}, \& {Cooray}}]{casey14}
{Casey}, C.~M., {Narayanan}, D., \& {Cooray}, A. 2014, \physrep, 541, 45

\bibitem[{{Chapin} {et~al.}(2009){Chapin}, {Pope}, {Scott}, {Aretxaga},
  {Austermann}, {Chary}, {Coppin}, {Halpern}, {Hughes}, {Lowenthal},
  {Morrison}, {Perera}, {Scott}, {Wilson}, \& {Yun}}]{chapin09}
{Chapin}, E.~L., {Pope}, A., {Scott}, D., {et~al.} 2009, \mnras, 398, 1793

\bibitem[{{Chapman} {et~al.}(2005){Chapman}, {Blain}, {Smail}, \&
  {Ivison}}]{chapman05}
{Chapman}, S.~C., {Blain}, A.~W., {Smail}, I., \& {Ivison}, R.~J. 2005, \apj,
  622, 772

\bibitem[{{Chen} {et~al.}(2016){Chen}, {Smail}, {Ivison}, {Arumugam},
  {Almaini}, {Conselice}, {Geach}, {Hartley}, {Ma}, {Mortlock}, {Simpson},
  {Simpson}, {Swinbank}, {Aretxaga}, {Blain}, {Chapman}, {Dunlop}, {Farrah},
  {Halpern}, {Micha{\l}owski}, {van der Werf}, {Wilkinson}, \&
  {Zavala}}]{chen16}
{Chen}, C.-C., {Smail}, I., {Ivison}, R.~J., {et~al.} 2016, \apj, 820, 82

\bibitem[{{da Cunha} {et~al.}(2015){da Cunha}, {Walter}, {Smail}, {Swinbank},
  {Simpson}, {Decarli}, {Hodge}, {Weiss}, {van der Werf}, {Bertoldi},
  {Chapman}, {Cox}, {Danielson}, {Dannerbauer}, {Greve}, {Ivison}, {Karim}, \&
  {Thomson}}]{dacunha15}
{da Cunha}, E., {Walter}, F., {Smail}, I.~R., {et~al.} 2015, \apj, 806, 110

\bibitem[{{Danielson} {et~al.}(2011){Danielson}, {Swinbank}, {Smail}, {Cox},
  {Edge}, {Weiss}, {Harris}, {Baker}, {De Breuck}, {Geach}, {Ivison}, {Krips},
  {Lundgren}, {Longmore}, {Neri}, \& {Flaquer}}]{danielson11}
{Danielson}, A.~L.~R., {Swinbank}, A.~M., {Smail}, I., {et~al.} 2011, \mnras,
  410, 1687

\bibitem[{{Dong} {et~al.}(2019){Dong}, {Spilker}, {Gonzalez}, {Apostolovski},
  {Aravena}, {B{\'e}thermin}, {Chapman}, {Chen}, {Hayward}, {Hezaveh}, {Litke},
  {Ma}, {Marrone}, {Morningstar}, {Phadke}, {Reuter}, {Sreevani}, {Stark},
  {Vieira}, \& {Wei{\ss}}}]{dong18}
{Dong}, C., {Spilker}, J.~S., {Gonzalez}, A.~H., {et~al.} 2019, arXiv e-prints
  [\eprint[arXiv]{1901.10482}]

\bibitem[{{Dye} {et~al.}(2008){Dye}, {Eales}, {Aretxaga}, {Serjeant}, {Dunlop},
  {Babbedge}, {Chapman}, {Cirasuolo}, {Clements}, {Coppin}, {Dunne}, {Egami},
  {Farrah}, {Ivison}, {van Kampen}, {Pope}, {Priddey}, {Rieke}, {Schael},
  {Scott}, {Simpson}, {Takagi}, {Takata}, \& {Vaccari}}]{dye08}
{Dye}, S., {Eales}, S.~A., {Aretxaga}, I., {et~al.} 2008, \mnras, 386, 1107

\bibitem[{{Fixsen} {et~al.}(1999){Fixsen}, {Bennett}, \& {Mather}}]{fixsen99}
{Fixsen}, D.~J., {Bennett}, C.~L., \& {Mather}, J.~C. 1999, \apj, 526, 207

\bibitem[{{Frayer} {et~al.}(1998){Frayer}, {Ivison}, {Scoville}, {Yun},
  {Evans}, {Smail}, {Blain}, \& {Kneib}}]{frayer98}
{Frayer}, D.~T., {Ivison}, R.~J., {Scoville}, N.~Z., {et~al.} 1998, \apjl, 506,
  L7

\bibitem[{{Gonz{\'a}lez-Alfonso} {et~al.}(2010){Gonz{\'a}lez-Alfonso},
  {Fischer}, {Isaak}, {Rykala}, {Savini}, {Spaans}, {van der Werf},
  {Meijerink}, {Israel}, {Loenen}, {Vlahakis}, {Smith}, {Charmandaris},
  {Aalto}, {Henkel}, {Wei{\ss}}, {Walter}, {Greve}, {Mart{\'{\i}}n-Pintado},
  {Naylor}, {Spinoglio}, {Veilleux}, {Harris}, {Armus}, {Lord}, {Mazzarella},
  {Xilouris}, {Sanders}, {Dasyra}, {Wiedner}, {Kramer}, {Papadopoulos},
  {Stacey}, {Evans}, \& {Gao}}]{gonzalezalfonso10}
{Gonz{\'a}lez-Alfonso}, E., {Fischer}, J., {Isaak}, K., {et~al.} 2010, \aap,
  518, L43

\bibitem[{{Greve} {et~al.}(2005){Greve}, {Bertoldi}, {Smail}, {Neri},
  {Chapman}, {Blain}, {Ivison}, {Genzel}, {Omont}, {Cox}, {Tacconi}, \&
  {Kneib}}]{greve05}
{Greve}, T.~R., {Bertoldi}, F., {Smail}, I., {et~al.} 2005, \mnras, 359, 1165

\bibitem[{{Gullberg} {et~al.}(2015){Gullberg}, {De Breuck}, {Vieira},
  {Wei{\ss}}, {Aguirre}, {Aravena}, {B{\'e}thermin}, {Bradford}, {Bothwell},
  {Carlstrom}, {Chapman}, {Fassnacht}, {Gonzalez}, {Greve}, {Hezaveh},
  {Holzapfel}, {Husband}, {Ma}, {Malkan}, {Marrone}, {Menten}, {Murphy},
  {Reichardt}, {Spilker}, {Stark}, {Strandet}, \& {Welikala}}]{gullberg15}
{Gullberg}, B., {De Breuck}, C., {Vieira}, J.~D., {et~al.} 2015, \mnras, 449,
  2883

\bibitem[{{G{\"u}sten} {et~al.}(2006){G{\"u}sten}, {Philipp}, {Wei{\ss}}, \&
  {Klein}}]{gusten06}
{G{\"u}sten}, R., {Philipp}, S.~D., {Wei{\ss}}, A., \& {Klein}, B. 2006, \aap,
  454, L115

\bibitem[{{Hainline} {et~al.}(2011){Hainline}, {Blain}, {Smail}, {Alexander},
  {Armus}, {Chapman}, \& {Ivison}}]{hainline11}
{Hainline}, L.~J., {Blain}, A.~W., {Smail}, I., {et~al.} 2011, \apj, 740, 96

\bibitem[{{Hezaveh} {et~al.}(2016){Hezaveh}, {Dalal}, {Marrone}, {Mao},
  {Morningstar}, {Wen}, {Blandford}, {Carlstrom}, {Fassnacht}, {Holder},
  {Kemball}, {Marshall}, {Murray}, {Perreault Levasseur}, {Vieira}, \&
  {Wechsler}}]{hezaveh16}
{Hezaveh}, Y.~D., {Dalal}, N., {Marrone}, D.~P., {et~al.} 2016, \apj, 823, 37

\bibitem[{{Hezaveh} {et~al.}(2013){Hezaveh}, {Marrone}, {Fassnacht}, {Spilker},
  {Vieira}, {Aguirre}, {Aird}, {Aravena}, {Ashby}, {Bayliss}, {Benson},
  {Bleem}, {Bothwell}, {Brodwin}, {Carlstrom}, {Chang}, {Chapman}, {Crawford},
  {Crites}, {De Breuck}, {de Haan}, {Dobbs}, {Fomalont}, {George}, {Gladders},
  {Gonzalez}, {Greve}, {Halverson}, {High}, {Holder}, {Holzapfel}, {Hoover},
  {Hrubes}, {Husband}, {Hunter}, {Keisler}, {Lee}, {Leitch}, {Lueker},
  {Luong-Van}, {Malkan}, {McIntyre}, {McMahon}, {Mehl}, {Menten}, {Meyer},
  {Mocanu}, {Murphy}, {Natoli}, {Padin}, {Plagge}, {Reichardt}, {Rest}, {Ruel},
  {Ruhl}, {Sharon}, {Schaffer}, {Shaw}, {Shirokoff}, {Stalder}, {Staniszewski},
  {Stark}, {Story}, {Vanderlinde}, {Wei{\ss}}, {Welikala}, \&
  {Williamson}}]{hezaveh13}
{Hezaveh}, Y.~D., {Marrone}, D.~P., {Fassnacht}, C.~D., {et~al.} 2013, \apj,
  767, 132

\bibitem[{{Hezaveh} {et~al.}(2012){Hezaveh}, {Marrone}, \&
  {Holder}}]{hezaveh12}
{Hezaveh}, Y.~D., {Marrone}, D.~P., \& {Holder}, G.~P. 2012, \apj, 761, 20

\bibitem[{{Hinshaw} {et~al.}(2013){Hinshaw}, {Larson}, {Komatsu}, {Spergel},
  {Bennett}, {Dunkley}, {Nolta}, {Halpern}, {Hill}, {Odegard}, {Page}, {Smith},
  {Weiland}, {Gold}, {Jarosik}, {Kogut}, {Limon}, {Meyer}, {Tucker}, {Wollack},
  \& {Wright}}]{hinshaw13}
{Hinshaw}, G., {Larson}, D., {Komatsu}, E., {et~al.} 2013, \apjs, 208, 19

\bibitem[{{Huynh} {et~al.}(2017){Huynh}, {Emonts}, {Kimball}, {Seymour},
  {Smail}, {Swinbank}, {Brandt}, {Casey}, {Chapman}, {Dannerbauer}, {Hodge},
  {Ivison}, {Schinnerer}, {Thomson}, {van der Werf}, \& {Wardlow}}]{huynh17}
{Huynh}, M.~T., {Emonts}, B.~H.~C., {Kimball}, A.~E., {et~al.} 2017, \mnras,
  467, 1222

\bibitem[{{Ivison} {et~al.}(2002){Ivison}, {Greve}, {Smail}, {Dunlop}, {Roche},
  {Scott}, {Page}, {Stevens}, {Almaini}, {Blain}, {Willott}, {Fox}, {Gilbank},
  {Serjeant}, \& {Hughes}}]{ivison02}
{Ivison}, R.~J., {Greve}, T.~R., {Smail}, I., {et~al.} 2002, \mnras, 337, 1

\bibitem[{{Kennicutt}(1998)}]{kennicutt98}
{Kennicutt}, Jr., R.~C. 1998, \araa, 36, 189

\bibitem[{{K{\"o}nig} {et~al.}(2017){K{\"o}nig}, {Mart{\'{\i}}n}, {Muller},
  {Cernicharo}, {Sakamoto}, {Zschaechner}, {Humphreys}, {Mroczkowski}, {Krips},
  {Galametz}, {Aalto}, {Vlemmings}, {Ott}, {Meier}, {Fuente},
  {Garc{\'{\i}}a-Burillo}, \& {Neri}}]{koenig17}
{K{\"o}nig}, S., {Mart{\'{\i}}n}, S., {Muller}, S., {et~al.} 2017, \aap, 602,
  A42

\bibitem[{{Litke} {et~al.}(2019){Litke}, {Marrone}, {Spilker}, {Aravena},
  {B{\'e}thermin}, {Chapman}, {Chen}, {de Breuck}, {Dong}, {Gonzalez}, {Greve},
  {Hayward}, {Hezaveh}, {Jarugula}, {Ma}, {Morningstar}, {Narayanan}, {Phadke},
  {Reuter}, {Vieira}, \& {Weiss}}]{litke18}
{Litke}, K.~C., {Marrone}, D.~P., {Spilker}, J.~S., {et~al.} 2019, \apj, 870,
  80

\bibitem[{{Liu} {et~al.}(2017){Liu}, {Wei{\ss}}, {Perez-Beaupuits},
  {G{\"u}sten}, {Liu}, {Gao}, {Menten}, {van der Werf}, {Israel}, {Harris},
  {Martin-Pintado}, {Requena-Torres}, \& {Stutzki}}]{liu17}
{Liu}, L., {Wei{\ss}}, A., {Perez-Beaupuits}, J.~P., {et~al.} 2017, \apj, 846,
  5

\bibitem[{{Ma} {et~al.}(2015){Ma}, {Gonzalez}, {Spilker}, {Strandet}, {Ashby},
  {Aravena}, {B{\'e}thermin}, {Bothwell}, {de Breuck}, {Brodwin}, {Chapman},
  {Fassnacht}, {Greve}, {Gullberg}, {Hezaveh}, {Malkan}, {Marrone},
  {Saliwanchik}, {Vieira}, {Weiss}, \& {Welikala}}]{ma15}
{Ma}, J., {Gonzalez}, A.~H., {Spilker}, J.~S., {et~al.} 2015, \apj, 812, 88

\bibitem[{{Ma} {et~al.}(2016){Ma}, {Gonzalez}, {Vieira}, {Aravena}, {Ashby},
  {B{\'e}thermin}, {Bothwell}, {Brandt}, {de Breuck}, {Carlstrom}, {Chapman},
  {Gullberg}, {Hezaveh}, {Litke}, {Malkan}, {Marrone}, {McDonald}, {Murphy},
  {Spilker}, {Sreevani}, {Stark}, {Strandet}, \& {Wang}}]{ma16}
{Ma}, J., {Gonzalez}, A.~H., {Vieira}, J.~D., {et~al.} 2016, \apj, 832, 114

\bibitem[{{Magnelli} {et~al.}(2012){Magnelli}, {Lutz}, {Santini}, {Saintonge},
  {Berta}, {Albrecht}, {Altieri}, {Andreani}, {Aussel}, {Bertoldi},
  {B{\'e}thermin}, {Bongiovanni}, {Capak}, {Chapman}, {Cepa}, {Cimatti},
  {Cooray}, {Daddi}, {Danielson}, {Dannerbauer}, {Dunlop}, {Elbaz}, {Farrah},
  {F{\"o}rster Schreiber}, {Genzel}, {Hwang}, {Ibar}, {Ivison}, {Le Floc'h},
  {Magdis}, {Maiolino}, {Nordon}, {Oliver}, {P{\'e}rez Garc{\'{\i}}a},
  {Poglitsch}, {Popesso}, {Pozzi}, {Riguccini}, {Rodighiero}, {Rosario},
  {Roseboom}, {Salvato}, {Sanchez-Portal}, {Scott}, {Smail}, {Sturm},
  {Swinbank}, {Tacconi}, {Valtchanov}, {Wang}, \& {Wuyts}}]{magnelli12}
{Magnelli}, B., {Lutz}, D., {Santini}, P., {et~al.} 2012, \aap, 539, A155

\bibitem[{{Marrone} {et~al.}(2018){Marrone}, {Spilker}, {Hayward}, {Vieira},
  {Aravena}, {Ashby}, {Bayliss}, {B{\'e}thermin}, {Brodwin}, {Bothwell},
  {Carlstrom}, {Chapman}, {Chen}, {Crawford}, {Cunningham}, {De Breuck},
  {Fassnacht}, {Gonzalez}, {Greve}, {Hezaveh}, {Lacaille}, {Litke}, {Lower},
  {Ma}, {Malkan}, {Miller}, {Morningstar}, {Murphy}, {Narayanan}, {Phadke},
  {Rotermund}, {Sreevani}, {Stalder}, {Stark}, {Strandet}, {Tang}, \&
  {Wei{\ss}}}]{marrone18}
{Marrone}, D.~P., {Spilker}, J.~S., {Hayward}, C.~C., {et~al.} 2018, \nat, 553,
  51

\bibitem[{{McMullin} {et~al.}(2007){McMullin}, {Waters}, {Schiebel}, {Young},
  \& {Golap}}]{mcmullin07}
{McMullin}, J.~P., {Waters}, B., {Schiebel}, D., {Young}, W., \& {Golap}, K.
  2007, in Astronomical Society of the Pacific Conference Series, Vol. 376,
  Astronomical Data Analysis Software and Systems XVI, ed. R.~A. {Shaw},
  F.~{Hill}, \& D.~J. {Bell}, 127

\bibitem[{{Micha{\l}owski} {et~al.}(2017){Micha{\l}owski}, {Dunlop},
  {Koprowski}, {Cirasuolo}, {Geach}, {Bowler}, {Mortlock}, {Caputi},
  {Aretxaga}, {Arumugam}, {Chen}, {McLure}, {Birkinshaw}, {Bourne}, {Farrah},
  {Ibar}, {van der Werf}, \& {Zemcov}}]{michalowski17}
{Micha{\l}owski}, M.~J., {Dunlop}, J.~S., {Koprowski}, M.~P., {et~al.} 2017,
  \mnras, 469, 492

\bibitem[{{Micha{\l}owski} {et~al.}(2014){Micha{\l}owski}, {Hayward}, {Dunlop},
  {Bruce}, {Cirasuolo}, {Cullen}, \& {Hernquist}}]{michalowski14}
{Micha{\l}owski}, M.~J., {Hayward}, C.~C., {Dunlop}, J.~S., {et~al.} 2014,
  \aap, 571, A75

\bibitem[{{Miettinen} {et~al.}(2017){Miettinen}, {Delvecchio}, {Smol{\v
  c}i{\'c}}, {Novak}, {Aravena}, {Karim}, {Murphy}, {Schinnerer}, {Capak},
  {Ilbert}, {Intema}, {Laigle}, \& {McCracken}}]{miettinen17}
{Miettinen}, O., {Delvecchio}, I., {Smol{\v c}i{\'c}}, V., {et~al.} 2017, \aap,
  597, A5

\bibitem[{{Omont}(2007)}]{omont07}
{Omont}, A. 2007, Reports on Progress in Physics, 70, 1099

\bibitem[{{Omont} {et~al.}(2013){Omont}, {Yang}, {Cox}, {Neri}, {Beelen},
  {Bussmann}, {Gavazzi}, {van der Werf}, {Riechers}, {Downes}, {Krips}, {Dye},
  {Ivison}, {Vieira}, {Weiss}, {Aguirre}, {Baes}, {Baker}, {Bertoldi},
  {Cooray}, {Dannerbauer}, {De Zotti}, {Eales}, {Fu}, {Gao}, {Guelin},
  {Harris}, {Jarvis}, {Lehnert}, {Leeuw}, {Lupu}, {Menten}, {Michalowski},
  {Negrello}, {Serjeant}, {Temi}, {Auld}, {Dariush}, {Dunne}, {Fritz},
  {Hopwood}, {Hoyos}, {Ibar}, {Maddox}, {Smith}, {Valiante}, {Bock},
  {Bradford}, {Glenn}, \& {Scott}}]{omont13}
{Omont}, A., {Yang}, C., {Cox}, P., {et~al.} 2013, ArXiv e-prints
  [\eprint[arXiv]{1301.6618}]

\bibitem[{{Rawle} {et~al.}(2014){Rawle}, {Egami}, {Bussmann}, {Gurwell},
  {Ivison}, {Boone}, {Combes}, {Danielson}, {Rex}, {Richard}, {Smail},
  {Swinbank}, {Altieri}, {Blain}, {Clement}, {Dessauges-Zavadsky}, {Edge},
  {Fazio}, {Jones}, {Kneib}, {Omont}, {P{\'e}rez-Gonz{\'a}lez}, {Schaerer},
  {Valtchanov}, {van der Werf}, {Walth}, {Zamojski}, \& {Zemcov}}]{rawle14}
{Rawle}, T.~D., {Egami}, E., {Bussmann}, R.~S., {et~al.} 2014, \apj, 783, 59

\bibitem[{{Riechers} {et~al.}(2013){Riechers}, {Bradford}, {Clements},
  {Dowell}, {P{\'e}rez-Fournon}, {Ivison}, {Bridge}, {Conley}, {Fu}, {Vieira},
  {Wardlow}, {Calanog}, {Cooray}, {Hurley}, {Neri}, {Kamenetzky}, {Aguirre},
  {Altieri}, {Arumugam}, {Benford}, {B{\'e}thermin}, {Bock}, {Burgarella},
  {Cabrera-Lavers}, {Chapman}, {Cox}, {Dunlop}, {Earle}, {Farrah}, {Ferrero},
  {Franceschini}, {Gavazzi}, {Glenn}, {Solares}, {Gurwell}, {Halpern},
  {Hatziminaoglou}, {Hyde}, {Ibar}, {Kov{\'a}cs}, {Krips}, {Lupu}, {Maloney},
  {Martinez-Navajas}, {Matsuhara}, {Murphy}, {Naylor}, {Nguyen}, {Oliver},
  {Omont}, {Page}, {Petitpas}, {Rangwala}, {Roseboom}, {Scott}, {Smith},
  {Staguhn}, {Streblyanska}, {Thomson}, {Valtchanov}, {Viero}, {Wang},
  {Zemcov}, \& {Zmuidzinas}}]{riechers13}
{Riechers}, D.~A., {Bradford}, C.~M., {Clements}, D.~L., {et~al.} 2013, \nat,
  496, 329

\bibitem[{{Riechers} {et~al.}(2006){Riechers}, {Walter}, {Carilli}, {Knudsen},
  {Lo}, {Benford}, {Staguhn}, {Hunter}, {Bertoldi}, {Henkel}, {Menten},
  {Weiss}, {Yun}, \& {Scoville}}]{riechers06}
{Riechers}, D.~A., {Walter}, F., {Carilli}, C.~L., {et~al.} 2006, \apj, 650,
  604

\bibitem[{{Rosenberg} {et~al.}(2015){Rosenberg}, {van der Werf}, {Aalto},
  {Armus}, {Charmandaris}, {D{\'{\i}}az-Santos}, {Evans}, {Fischer}, {Gao},
  {Gonz{\'a}lez-Alfonso}, {Greve}, {Harris}, {Henkel}, {Israel}, {Isaak},
  {Kramer}, {Meijerink}, {Naylor}, {Sanders}, {Smith}, {Spaans}, {Spinoglio},
  {Stacey}, {Veenendaal}, {Veilleux}, {Walter}, {Wei{\ss}}, {Wiedner}, {van der
  Wiel}, \& {Xilouris}}]{rosenberg15}
{Rosenberg}, M.~J.~F., {van der Werf}, P.~P., {Aalto}, S., {et~al.} 2015, \apj,
  801, 72

\bibitem[{{Sargent} {et~al.}(2014){Sargent}, {Daddi}, {B{\'e}thermin},
  {Aussel}, {Magdis}, {Hwang}, {Juneau}, {Elbaz}, \& {da Cunha}}]{sargent14}
{Sargent}, M.~T., {Daddi}, E., {B{\'e}thermin}, M., {et~al.} 2014, \apj, 793,
  19

\bibitem[{{Serjeant}(2012)}]{serjeant12}
{Serjeant}, S. 2012, \mnras, 424, 2429

\bibitem[{{Sharon}(2013)}]{sharon13}
{Sharon}, C.~E. 2013, PhD thesis, Rutgers The State University of New Jersey -
  New Brunswick

\bibitem[{{Simpson} {et~al.}(2014){Simpson}, {Swinbank}, {Smail}, {Alexander},
  {Brandt}, {Bertoldi}, {de Breuck}, {Chapman}, {Coppin}, {da Cunha},
  {Danielson}, {Dannerbauer}, {Greve}, {Hodge}, {Ivison}, {Karim}, {Knudsen},
  {Poggianti}, {Schinnerer}, {Thomson}, {Walter}, {Wardlow}, {Wei{\ss}}, \&
  {van der Werf}}]{simpson14}
{Simpson}, J.~M., {Swinbank}, A.~M., {Smail}, I., {et~al.} 2014, \apj, 788, 125

\bibitem[{{Smolcic} {et~al.}(2012){Smolcic}, {Aravena}, {Navarrete},
  {Schinnerer}, {Riechers}, {Bertoldi}, {Feruglio}, {Finoguenov}, {Salvato},
  {Sargent}, {McCracken}, {Albrecht}, {Karim}, {Capak}, {Carilli},
  {Cappelluti}, {Elvis}, {Ilbert}, {Kartaltepe}, {Lilly}, {Sanders}, {Sheth},
  {Scoville}, \& {Taniguchi}}]{smolcic12}
{Smolcic}, V., {Aravena}, M., {Navarrete}, F., {et~al.} 2012, ArXiv e-prints
  [\eprint[arXiv]{1205.6470}]

\bibitem[{{Solomon} \& {Vanden Bout}(2005)}]{solomon05}
{Solomon}, P.~M. \& {Vanden Bout}, P.~A. 2005, \araa, 43, 677

\bibitem[{{Spilker} {et~al.}(2015){Spilker}, {Aravena}, {Marrone},
  {B{\'e}thermin}, {Bothwell}, {Carlstrom}, {Chapman}, {Collier}, {de Breuck},
  {Fassnacht}, {Galvin}, {Gonzalez}, {Gonz{\'a}lez-L{\'o}pez}, {Grieve},
  {Hezaveh}, {Ma}, {Malkan}, {O'Brien}, {Rotermund}, {Strandet}, {Vieira},
  {Weiss}, \& {Wong}}]{spilker15}
{Spilker}, J.~S., {Aravena}, M., {Marrone}, D.~P., {et~al.} 2015, \apj, 811,
  124

\bibitem[{{Spilker} {et~al.}(2014){Spilker}, {Marrone}, {Aguirre}, {Aravena},
  {Ashby}, {B{\'e}thermin}, {Bradford}, {Bothwell}, {Brodwin}, {Carlstrom},
  {Chapman}, {Crawford}, {de Breuck}, {Fassnacht}, {Gonzalez}, {Greve},
  {Gullberg}, {Hezaveh}, {Holzapfel}, {Husband}, {Ma}, {Malkan}, {Murphy},
  {Reichardt}, {Rotermund}, {Stalder}, {Stark}, {Strandet}, {Vieira},
  {Wei{\ss}}, \& {Welikala}}]{spilker14}
{Spilker}, J.~S., {Marrone}, D.~P., {Aguirre}, J.~E., {et~al.} 2014, \apj, 785,
  149

\bibitem[{{Spilker} {et~al.}(2016){Spilker}, {Marrone}, {Aravena},
  {B{\'e}thermin}, {Bothwell}, {Carlstrom}, {Chapman}, {Crawford}, {de Breuck},
  {Fassnacht}, {Gonzalez}, {Greve}, {Hezaveh}, {Litke}, {Ma}, {Malkan},
  {Rotermund}, {Strandet}, {Vieira}, {Weiss}, \& {Welikala}}]{spilker16}
{Spilker}, J.~S., {Marrone}, D.~P., {Aravena}, M., {et~al.} 2016, \apj, 826,
  112

\bibitem[{{Strandet} {et~al.}(2017){Strandet}, {Weiss}, {De Breuck}, {Marrone},
  {Vieira}, {Aravena}, {Ashby}, {B{\'e}thermin}, {Bothwell}, {Bradford},
  {Carlstrom}, {Chapman}, {Cunningham}, {Chen}, {Fassnacht}, {Gonzalez},
  {Greve}, {Gullberg}, {Hayward}, {Hezaveh}, {Litke}, {Ma}, {Malkan}, {Menten},
  {Miller}, {Murphy}, {Narayanan}, {Phadke}, {Rotermund}, {Spilker}, \&
  {Sreevani}}]{strandet17}
{Strandet}, M.~L., {Weiss}, A., {De Breuck}, C., {et~al.} 2017, \apjl, 842, L15

\bibitem[{{Strandet} {et~al.}(2016){Strandet}, {Weiss}, {Vieira}, {de Breuck},
  {Aguirre}, {Aravena}, {Ashby}, {B{\'e}thermin}, {Bradford}, {Carlstrom},
  {Chapman}, {Crawford}, {Everett}, {Fassnacht}, {Furstenau}, {Gonzalez},
  {Greve}, {Gullberg}, {Hezaveh}, {Kamenetzky}, {Litke}, {Ma}, {Malkan},
  {Marrone}, {Menten}, {Murphy}, {Nadolski}, {Rotermund}, {Spilker}, {Stark},
  \& {Welikala}}]{strandet16}
{Strandet}, M.~L., {Weiss}, A., {Vieira}, J.~D., {et~al.} 2016, \apj, 822, 80

\bibitem[{{Swinbank} {et~al.}(2014){Swinbank}, {Simpson}, {Smail}, {Harrison},
  {Hodge}, {Karim}, {Walter}, {Alexander}, {Brandt}, {de Breuck}, {da Cunha},
  {Chapman}, {Coppin}, {Danielson}, {Dannerbauer}, {Decarli}, {Greve},
  {Ivison}, {Knudsen}, {Lagos}, {Schinnerer}, {Thomson}, {Wardlow}, {Wei{\ss}},
  \& {van der Werf}}]{swinbank14}
{Swinbank}, A.~M., {Simpson}, J.~M., {Smail}, I., {et~al.} 2014, \mnras, 438,
  1267

\bibitem[{{van der Werf} {et~al.}(2011){van der Werf}, {Berciano Alba},
  {Spaans}, {Loenen}, {Meijerink}, {Riechers}, {Cox}, {Wei{\ss}}, \&
  {Walter}}]{vanderwerf11}
{van der Werf}, P.~P., {Berciano Alba}, A., {Spaans}, M., {et~al.} 2011, \apjl,
  741, L38

\bibitem[{{Vieira} {et~al.}(2010){Vieira}, {Crawford}, {Switzer}, {Ade},
  {Aird}, {Ashby}, {Benson}, {Bleem}, {Brodwin}, {Carlstrom}, {Chang}, {Cho},
  {Crites}, {de Haan}, {Dobbs}, {Everett}, {George}, {Gladders}, {Hall},
  {Halverson}, {High}, {Holder}, {Holzapfel}, {Hrubes}, {Joy}, {Keisler},
  {Knox}, {Lee}, {Leitch}, {Lueker}, {Marrone}, {McIntyre}, {McMahon}, {Mehl},
  {Meyer}, {Mohr}, {Montroy}, {Padin}, {Plagge}, {Pryke}, {Reichardt}, {Ruhl},
  {Schaffer}, {Shaw}, {Shirokoff}, {Spieler}, {Stalder}, {Staniszewski},
  {Stark}, {Vanderlinde}, {Walsh}, {Williamson}, {Yang}, {Zahn}, \&
  {Zenteno}}]{vieira10}
{Vieira}, J.~D., {Crawford}, T.~M., {Switzer}, E.~R., {et~al.} 2010, \apj, 719,
  763

\bibitem[{{Vieira} {et~al.}(2013){Vieira}, {Marrone}, {Chapman}, {De Breuck},
  {Hezaveh}, {Wei{$\beta$}}, {Aguirre}, {Aird}, {Aravena}, {Ashby}, {Bayliss},
  {Benson}, {Biggs}, {Bleem}, {Bock}, {Bothwell}, {Bradford}, {Brodwin},
  {Carlstrom}, {Chang}, {Crawford}, {Crites}, {de Haan}, {Dobbs}, {Fomalont},
  {Fassnacht}, {George}, {Gladders}, {Gonzalez}, {Greve}, {Gullberg},
  {Halverson}, {High}, {Holder}, {Holzapfel}, {Hoover}, {Hrubes}, {Hunter},
  {Keisler}, {Lee}, {Leitch}, {Lueker}, {Luong-van}, {Malkan}, {McIntyre},
  {McMahon}, {Mehl}, {Menten}, {Meyer}, {Mocanu}, {Murphy}, {Natoli}, {Padin},
  {Plagge}, {Reichardt}, {Rest}, {Ruel}, {Ruhl}, {Sharon}, {Schaffer}, {Shaw},
  {Shirokoff}, {Spilker}, {Stalder}, {Staniszewski}, {Stark}, {Story},
  {Vanderlinde}, {Welikala}, \& {Williamson}}]{vieira13}
{Vieira}, J.~D., {Marrone}, D.~P., {Chapman}, S.~C., {et~al.} 2013, \nat, 495,
  344

\bibitem[{{Wardlow} {et~al.}(2011){Wardlow}, {Smail}, {Coppin}, {Alexander},
  {Brandt}, {Danielson}, {Luo}, {Swinbank}, {Walter}, {Wei{\ss}}, {Xue},
  {Zibetti}, {Bertoldi}, {Biggs}, {Chapman}, {Dannerbauer}, {Dunlop},
  {Gawiser}, {Ivison}, {Knudsen}, {Kov{\'a}cs}, {Lacey}, {Menten}, {Padilla},
  {Rix}, \& {van der Werf}}]{wardlow11}
{Wardlow}, J.~L., {Smail}, I., {Coppin}, K.~E.~K., {et~al.} 2011, \mnras, 415,
  1479

\bibitem[{{Wei{\ss}} {et~al.}(2013){Wei{\ss}}, {De Breuck}, {Marrone},
  {Vieira}, {Aguirre}, {Aird}, {Aravena}, {Ashby}, {Bayliss}, {Benson},
  {B{\'e}thermin}, {Biggs}, {Bleem}, {Bock}, {Bothwell}, {Bradford}, {Brodwin},
  {Carlstrom}, {Chang}, {Chapman}, {Crawford}, {Crites}, {de Haan}, {Dobbs},
  {Downes}, {Fassnacht}, {George}, {Gladders}, {Gonzalez}, {Greve},
  {Halverson}, {Hezaveh}, {High}, {Holder}, {Holzapfel}, {Hoover}, {Hrubes},
  {Husband}, {Keisler}, {Lee}, {Leitch}, {Lueker}, {Luong-Van}, {Malkan},
  {McIntyre}, {McMahon}, {Mehl}, {Menten}, {Meyer}, {Murphy}, {Padin},
  {Plagge}, {Reichardt}, {Rest}, {Rosenman}, {Ruel}, {Ruhl}, {Schaffer},
  {Shirokoff}, {Spilker}, {Stalder}, {Staniszewski}, {Stark}, {Story},
  {Vanderlinde}, {Welikala}, \& {Williamson}}]{weiss13}
{Wei{\ss}}, A., {De Breuck}, C., {Marrone}, D.~P., {et~al.} 2013, \apj, 767, 88

\bibitem[{{Wei{\ss}} {et~al.}(2007){Wei{\ss}}, {Downes}, {Neri}, {Walter},
  {Henkel}, {Wilner}, {Wagg}, \& {Wiklind}}]{weiss07}
{Wei{\ss}}, A., {Downes}, D., {Neri}, R., {et~al.} 2007, \aap, 467, 955

\bibitem[{{Wei{\ss}} {et~al.}(2005{\natexlab{a}}){Wei{\ss}}, {Downes},
  {Walter}, \& {Henkel}}]{weiss05b}
{Wei{\ss}}, A., {Downes}, D., {Walter}, F., \& {Henkel}, C. 2005{\natexlab{a}},
  \aap, 440, L45

\bibitem[{{Wei{\ss}} {et~al.}(2005{\natexlab{b}}){Wei{\ss}}, {Walter}, \&
  {Scoville}}]{weiss05a}
{Wei{\ss}}, A., {Walter}, F., \& {Scoville}, N.~Z. 2005{\natexlab{b}}, \aap,
  438, 533

\bibitem[{{Yang} {et~al.}(2013){Yang}, {Gao}, {Omont}, {Liu}, {Isaak},
  {Downes}, {van der Werf}, \& {Lu}}]{yang13}
{Yang}, C., {Gao}, Y., {Omont}, A., {et~al.} 2013, \apjl, 771, L24

\bibitem[{{Yang} {et~al.}(2016){Yang}, {Omont}, {Beelen},
  {Gonz{\'a}lez-Alfonso}, {Neri}, {Gao}, {van der Werf}, {Wei{\ss}}, {Gavazzi},
  {Falstad}, {Baker}, {Bussmann}, {Cooray}, {Cox}, {Dannerbauer}, {Dye},
  {Gu{\'e}lin}, {Ivison}, {Krips}, {Lehnert}, {Micha{\l}owski}, {Riechers},
  {Spaans}, \& {Valiante}}]{yang16}
{Yang}, C., {Omont}, A., {Beelen}, A., {et~al.} 2016, \aap, 595, A80

\bibitem[{{Yun} {et~al.}(2012){Yun}, {Scott}, {Guo}, {Aretxaga}, {Giavalisco},
  {Austermann}, {Capak}, {Chen}, {Ezawa}, {Hatsukade}, {Hughes}, {Iono},
  {Johnson}, {Kawabe}, {Kohno}, {Lowenthal}, {Miller}, {Morrison}, {Oshima},
  {Perera}, {Salvato}, {Silverman}, {Tamura}, {Williams}, \& {Wilson}}]{yun12}
{Yun}, M.~S., {Scott}, K.~S., {Guo}, Y., {et~al.} 2012, \mnras, 420, 957

\end{thebibliography}
\onecolumn

\begin{appendix}
\section{}

\begin{table}[tbh!]
\centering
\caption{H$_2$O Source Reconstruction Results \label{tab:sourcemodelH2O}}
\begin{tabular}{ccccccc}
\hline \hline
{Line} & {Velocity} & {$\mu_{\text{H$_2$O}}$} & {$r_{\text{eff,H$_2$O}}$} & {I$_{\text{int,H$_2$O}}$} &  {$L'_{\text{int,H$_2$O}}$} & {SNR} \\
{} & {(km s$^{-1}$)} & {} & {(kpc)} & {(mJy km s$^{-1}$)} & {($10^{9}$ K km s$^{-1}$ pc$^{2}$)}\\
\hline
H$_2$O $2_{11}-2_{02}$	& $ -650 $ 	& 5.48 $\pm$ 1.29 & 1.65 $\pm$ 0.62 &  158.3 $\pm$ 45.7 &  4.1$\pm$ 1.2 & 5.98 \\
 						& $ -400 $ 	& 3.64 $\pm$ 0.45 & 2.55 $\pm$ 0.58 & 118.2 $\pm$ 24.1 & 3.1 $\pm$ 0.6 &   6.2    \\
 						& $ -300 $ 	& 4.70 $\pm$ 0.50 & 1.53 $\pm$ 0.39 & 112.1 $\pm$ 19.0 & 2.9 $\pm$ 0.5 &   7.5    \\
 						& $ -200 $ 	& 5.34 $\pm$ 0.63 & 1.79 $\pm$ 0.39 & 143.7 $\pm$ 21.3 & 3.7 $\pm$ 0.5 &   11.0   \\
 						& $ -100 $ 	& 7.89 $\pm$ 0.73 & 1.14 $\pm$ 0.16 & 104.5 $\pm$ 13.1 & 2.7 $\pm$ 0.3 &   11.8    \\
						& $ 0 $ 	& 8.32 $\pm$ 0.76 & 0.95 $\pm$ 0.14 & 100.0 $\pm$ 12.4 & 2.6 $\pm$ 0.3 &   12.0    \\
						& $ 100 $ 	& 9.32 $\pm$ 0.52 & 0.86 $\pm$ 0.10 & 111.3 $\pm$ 9.7 & 2.9 $\pm$ 0.2 &   14.8   \\
						& $ 200 $ 	& 9.16 $\pm$ 0.57 & 0.83 $\pm$ 0.10 & 99.2 $\pm$ 9.8 & 2.6 $\pm$ 0.2 &   13.0   \\
						& $ 300 $ 	& 8.94 $\pm$ 0.88 & 0.97 $\pm$ 0.23 & 50.0 $\pm$ 9.2 & 1.3 $\pm$ 0.2 &   6.4    \\
						& $ 400 $ 	& 6.20 $\pm$ 1.32 & 2.09 $\pm$ 0.67 & 69.5 $\pm$ 18.6 & 1.8 $\pm$ 0.4 &   6.2   \\
						& $ 650 $ 	& 6.09 $\pm$ 1.28 & 1.40 $\pm$ 0.43 &  95.7 $\pm$ 31.2 &  2.5$ \pm$ 0.8 & 4.0 \\
H$_2$O $2_{02}-1_{11}$ 	& $ -650 $ 	& 4.76 $\pm$ 0.60 & 0.84 $\pm$ 0.05 & 118.1 $\pm$ 29.8  & 1.8  $\pm$ 0.4 & 4.7 \\
						& $ -400 $ 	& 4.45 $\pm$ 0.45 & 0.82 $\pm$ 0.05 & 127.2 $\pm$ 18.5 & 1.9 $\pm$ 0.2 &   9.4    \\ 
 						& $ -300 $ 	& 4.53 $\pm$ 0.05 & 0.78 $\pm$ 0.04 & 166.7 $\pm$ 13.4 & 2.5 $\pm$ 0.2 &   12.6   \\
 						& $ -200 $ 	& 5.33 $\pm$ 0.34 & 1.05 $\pm$ 0.07 & 178.3 $\pm$ 16.1 & 2.7 $\pm$ 0.2 &   15.8   \\
 						& $ -100 $ 	& 5.84 $\pm$ 0.08 & 1.12 $\pm$ 0.07 & 171.2 $\pm$ 10.5 & 2.6 $\pm$ 0.1 &   16.7  \\
						& $ 0 $ 	& 6.11 $\pm$ 0.27 & 0.94 $\pm$ 0.07 & 182.3 $\pm$ 12.6 & 2.7 $\pm$ 0.1 &   18.6   \\
						& $ 100 $ 	& 6.49 $\pm$ 0.08 & 0.98 $\pm$ 0.05 & 230.0 $\pm$ 9.6 & 3.5 $\pm$ 0.1 &   24.9   \\
						& $ 200 $ 	& 6.22 $\pm$ 0.19 & 0.86 $\pm$ 0.06 & 216.0 $\pm$ 11.7 & 3.2 $\pm$ 0.1 &   22.4   \\
						& $ 300 $ 	& 5.71 $\pm$ 0.30 & 0.82 $\pm$ 0.05 & 145.2 $\pm$ 13.0 & 2.2 $\pm$ 0.1 &   13.8   \\
						& $ 400 $ 	& 5.54 $\pm$ 0.23 & 0.78 $\pm$ 0.06 & 71.3 $\pm$ 11.2 & 1.0 $\pm$ 0.1 &   6.6    \\
						& $ 650 $ 	& 5.18 $\pm$ 0.23 & 0.82 $\pm$ 0.05 &  66.3 $\pm$ 24.0 &  1.0$\pm$ 0.3 &2.8\\
\hline
\end{tabular}
\end{table}

\begin{table}[tbh!]
\centering
\caption{CO Source Reconstruction Results \label{tab:sourcemodelCO}}
\begin{tabular}{ccccccc}
\hline \hline
{Line} & {Velocity} & {$\mu_{\text{CO}}$} & {$r_{\text{eff,CO}}$} & {I$_{\text{CO}}$} & {$L'_{\text{int,CO}}$} & {SNR} \\
{} & {(km s$^{-1}$)} & {} & {(kpc)} & {(mJy km s$^{-1}$)} & {($10^{9}$ K km s$^{-1}$ pc$^{2}$)}\\
\hline
CO(2$-$1) 	& $ -400 $ 	& 5.36 $\pm$ 1.00	& 1.81 $\pm$ 0.75 	& 26.2 $\pm$ 7.4    & 7.3 $\pm$ 2.0 & 4.7    \\
 			& $ -300 $ 	& 5.46 $\pm$ 0.91 	& 2.66 $\pm$ 0.53 	& 33.8 $\pm$ 7.8 	& 9.4 $\pm$ 2.2 & 6.1    \\
 			& $ -200 $ 	& 4.82 $\pm$ 0.75 	& 1.90 $\pm$ 0.41 	& 46.9 $\pm$ 9.6 	& 13.1 $\pm$ 2.6 & 7.5    \\
 			& $ -100 $ 	& 6.27 $\pm$ 1.12 	& 1.94 $\pm$ 0.56 	& 40.9 $\pm$ 8.7 	& 11.4 $\pm$ 2.4 & 8.6    \\
 			& $ 0 $ 	& 9.15 $\pm$ 0.83 	& 1.12 $\pm$ 0.27 	& 32.8 $\pm$ 4.4 	& 9.2 $\pm$ 1.2 & 10.0    \\
 			& $ 100 $ 	& 8.82 $\pm$ 0.59 	& 1.02 $\pm$ 0.12 	& 32.4 $\pm$ 4.0 	& 9.0 $\pm$ 1.1 & 9.5    \\
			& $ 200 $ 	& 9.59 $\pm$ 0.69 	& 0.86 $\pm$ 0.15 	& 38.7 $\pm$ 4.1 	& 10.8 $\pm$ 1.1 & 12.4    \\
			& $ 300 $ 	& 6.95 $\pm$ 2.08 	& 1.74 $\pm$ 0.82 	& 28.3 $\pm$ 9.5 	& 7.9 $\pm$ 2.6 & 6.6    \\
			& $ 400 $ 	& 7.16 $\pm$ 3.38 	& 0.85 $\pm$ 0.80 	& 6.7  $\pm$ 5.2 	& 1.8 $\pm$ 1.4 & 1.6    \\
CO(6$-$5) 	& $ -650 $ 	& 3.64 $\pm$ 0.10 	& 0.79 $\pm$ 0.06 	& 218.6 $\pm$ 46.9 	& 6.8 $\pm$ 1.4 & 4.6     \\
			& $ -400 $ 	& 3.68 $\pm$ 0.79 	& 0.83 $\pm$ 0.06 	& 177.5 $\pm$ 43.7 	& 5.5 $\pm$ 1.3 & 8.2    \\
 			& $ -300 $ 	& 4.47 $\pm$ 0.09 	& 0.95 $\pm$ 0.05 	& 236.3 $\pm$ 18.5 	& 7.3 $\pm$ 0.5 & 13.2   \\
 			& $ -200 $ 	& 5.72 $\pm$ 0.12 	& 0.97 $\pm$ 0.05 	& 241.0 $\pm$ 14.9 	& 7.5 $\pm$ 0.4 & 17.2   \\
 			& $ -100 $ 	& 7.73 $\pm$ 0.23 	& 0.98 $\pm$ 0.05 	& 192.7 $\pm$ 11.8 	& 6.0 $\pm$ 0.3 & 18.6   \\
			& $ 0 $ 	& 8.38 $\pm$ 0.21 	& 0.82 $\pm$ 0.03 	& 181.4 $\pm$ 10.5 	& 5.6 $\pm$ 0.3 & 19.0   \\
			& $ 100 $ 	& 8.80 $\pm$ 0.19 	& 0.76 $\pm$ 0.02 	& 209.8 $\pm$ 10.1 	& 6.5 $\pm$ 0.3 & 23.1   \\
			& $ 200 $ 	& 8.24 $\pm$ 0.16 	& 0.66 $\pm$ 0.02 	& 190.5 $\pm$ 10.4 	& 5.9 $\pm$ 0.3 & 19.6   \\
			& $ 300 $ 	& 8.13 $\pm$ 0.29 	& 0.67 $\pm$ 0.05 	& 114.7 $\pm$ 10.6 	& 3.5 $\pm$ 0.3 & 11.7   \\
			& $ 400 $ 	& 8.73 $\pm$ 0.68 	& 0.53 $\pm$ 0.10 	&  48.2 $\pm$ 9.8   & 1.5 $\pm$ 0.3 & 5.3    \\
			& $ 650 $ 	& 7.40 $\pm$ 0.46 	& 0.30 $\pm$ 0.09 	& 96.2 $\pm$ 23.8  	& 2.9 $\pm$ 0.7 & 4.17\\
CO(8$-$7) 	& $ -650 $ 	& 4.57 $\pm$ 0.64 	& 0.77 $\pm$ 0.07 	& 147.8 $\pm$  31.9	& 2.5 $\pm$ 0.5 & 6.07 \\
			& $ -400 $ 	& 4.42 $\pm$ 1.11 	& 0.61 $\pm$ 0.07	& 139.6 $\pm$ 37.0 	& 2.4 $\pm$ 0.5 & 12.3   \\
 			& $ -300 $ 	& 4.72 $\pm$ 0.43 	& 0.84 $\pm$ 0.07 	& 201.5 $\pm$ 21.1 	& 3.5 $\pm$ 0.3 & 19.0    \\
 			& $ -200 $ 	& 5.64 $\pm$ 0.08 	& 1.00 $\pm$ 0.05 	& 220.0 $\pm$ 9.3 	& 3.8 $\pm$ 0.1 & 24.8   \\
			& $ -100 $ 	& 6.02 $\pm$ 0.08 	& 1.09 $\pm$ 0.07 	& 228.9 $\pm$ 8.8 	& 4.0 $\pm$ 0.1 & 27.6   \\
			& $ 0 $ 	& 6.29 $\pm$ 0.08 	& 1.10 $\pm$ 0.08 	& 224.7 $\pm$ 8.4 	& 3.9 $\pm$ 0.1 & 28.3  \\
			& $ 100 $ 	& 6.72 $\pm$ 0.09 	& 0.90 $\pm$ 0.04 	& 258.4 $\pm$ 8.1 	& 4.5 $\pm$ 0.1 & 34.8   \\
			& $ 200 $ 	& 6.69 $\pm$ 0.18 	& 0.86 $\pm$ 0.06 	& 208.7 $\pm$ 9.3 	& 3.6 $\pm$ 0.1 & 27.9   \\
			& $ 300 $ 	& 5.81 $\pm$ 0.12 	& 1.10 $\pm$ 0.11 	& 140.5 $\pm$ 9.0 	& 2.4 $\pm$ 0.1 & 16.3   \\
			& $ 400 $ 	& 5.24 $\pm$ 0.09 	& 0.70 $\pm$ 0.08 	&  77.3 $\pm$ 9.6 	& 1.3 $\pm$ 0.1 & 8.1    \\
			& $ 650 $ 	& 5.14 $\pm$ 0.34 	& 0.80 $\pm$ 0.05 	&  35.6 $\pm$ 21.7 	& 0.6  $\pm$0.3  & 1.6\\
CO(9$-$8) 	& $ -650 $ 	& 4.70 $\pm$ 0.42 	& 0.71 $\pm$ 0.05 	&  88.0 $\pm$ 76.1	& 1.2  $\pm$ 1.0 & 1.7 \\
			& $ -400 $ 	& 4.64 $\pm$ 0.76 	& 0.73 $\pm$ 0.09 	& 198.1 $\pm$ 45.8	& 2.7 $\pm$ 0.6 & 6.1    \\
			& $ -300 $ 	& 4.81 $\pm$ 0.29 	& 0.74 $\pm$ 0.06 	& 168.6 $\pm$ 32.8 	& 2.3 $\pm$ 0.4 & 5.4    \\
 			& $ -200 $ 	& 5.45 $\pm$ 0.22 	& 0.84 $\pm$ 0.07 	& 234.5 $\pm$ 29.0	& 3.2 $\pm$ 0.4 & 8.5    \\
 			& $ -100 $ 	& 6.20 $\pm$ 0.13 	& 0.85 $\pm$ 0.07 	& 202.6 $\pm$ 24.6 	& 2.8 $\pm$ 0.3 & 8.4    \\
			& $ 0 $ 	& 5.49 $\pm$ 0.04 	& 0.77 $\pm$ 0.03 	& 245.9 $\pm$ 27.3 	& 3.4 $\pm$ 0.3 & 9.0    \\
			& $ 100 $ 	& 6.34 $\pm$ 0.15 	& 0.81 $\pm$ 0.07 	& 239.4 $\pm$ 24.2 	& 3.3 $\pm$ 0.3 & 10.1    \\
			& $ 200 $ 	& 6.61 $\pm$ 0.18 	& 0.74 $\pm$ 0.08 	& 249.3 $\pm$ 23.7 	& 3.4 $\pm$ 0.3 & 11.0    \\
			& $ 300 $ 	& 6.17 $\pm$ 0.19 	& 1.0 $\pm$ 0.14 	& 165.1 $\pm$ 24.8	& 2.2 $\pm$ 0.3 & 6.8    \\
			& $ 400 $ 	& 5.17 $\pm$ 0.57 	& 0.92 $\pm$ 0.14 	&  68.4 $\pm$ 29.9 	& 1.0 $\pm$ 0.4 & 2.4    \\
			& $ 650 $ 	& 5.32 $\pm$ 0.07 	& 0.81 $\pm$ 0.05 	& 165.7 $\pm$ 67.2  & 2.2 $\pm$ 0.9 & 2.5\\
\hline			
\end{tabular}
\end{table}

\end{appendix}

\end{document}